\documentclass[aps,prd,superscriptaddress,showpacs,preprint,amsmath,amssymb]{revtex4}
\usepackage{graphicx, bm}
\usepackage[usenames]{color}

\usepackage{subcaption}
\begin{document}

%\tightenlines
\draft
\title{Future projections on the anomalous $WW\gamma\gamma$ couplings in hadron-hadron interactions at the FCC-hh}

\author{ A. Guti\'errez-Rodr\'{\i}guez\footnote{alexgu@fisica.uaz.edu.mx}}
\affiliation{\small Facultad de F\'{\i}sica, Universidad Aut\'onoma de Zacatecas\\
         Apartado Postal C-580, 98060 Zacatecas, M\'exico.\\}

\author{V. Ari\footnote{vari@science.ankara.edu.tr}}
\affiliation{\small Department of Physics, Ankara University, Turkey.\\}

\author{E. Gurkanli\footnote{egurkanli@sinop.edu.tr}}
\affiliation{\small Department of Physics, Sinop University, Turkey.\\}

\author{M. K\"{o}ksal\footnote{mkoksal@cumhuriyet.edu.tr}}
\affiliation{\small Department of Physics, Sivas Cumhuriyet University, 58140, Sivas, Turkey.}

\author{ M. A. Hern\'andez-Ru\'{\i}z\footnote{mahernan@uaz.edu.mx}}
\affiliation{\small Unidad Acad\'emica de Ciencias Qu\'{\i}micas, Universidad Aut\'onoma de Zacatecas\\
         Apartado Postal C-585, 98060 Zacatecas, M\'exico.\\}

\date{\today}
%\maketitle

\begin{abstract}
% insert abstract here

We carry out a study in the approach of a dimension-8 effective field theory on the anomalous quartic $WW\gamma\gamma$
couplings that arise from the process $pp \to W\gamma\gamma$ at the Future Circular Collider-hadron hadron (FCC-hh).
The process is sensitive to the $WW\gamma\gamma$ vertex, and future projections on the anomalous $f_ {M,i}/\Lambda^4$ and
$ f_ {T,j}/\Lambda^4$ couplings can be put by measurements of $W\gamma\gamma$ production. We illustrate the physics opportunities
and reach of the FCC-hh operating for a high integrated luminosity scenario of ${\cal L}=1, 5, 10, 30$ ${\rm ab}^{-1}$ of
proton-proton collisions at a center-of-mass energy of $\sqrt{s}=100$ TeV and systematic uncertainties of $\delta_{sys}=0\%,
3\%, 5\%, 10\%$. Production cross-sections of the signal and the sensitivity constraints on the anomalous $f_ {M,i}/\Lambda^4$
and $ f_ {T,j}/\Lambda^4$ couplings at $95\%$ C.L. are identified through the leptonic and hadronic decay modes of the $W$-boson.
The FCC-hh sensitivity on these anomalous couplings is expected might reach up to the order of magnitude ${\cal O}(10^{-4}-10^1)$.
These sensitivity perspectives indicate that the $pp \to W\gamma\gamma$ channel at the FCC-hh has a greater potential to search
for the anomalous quartic $WW\gamma\gamma$ couplings and to search for new physics than present colliders.

\end{abstract}


\vspace{5mm}

\maketitle

\section{Introduction}

The theoretical, phenomenological, and experimental research of the physics of elementary particles at the energy
frontier has entered an exciting era. In this regard, the experiments at the Large Hadron Collider (LHC) and future
collider experiments for the post-LHC era, such as the High-Energy LHC (HE-LHC) \cite{HL-LHC-HE-LHC},
High-Luminosity LHC (HL-LHC) \cite{HL-LHC-HE-LHC} and the Future Circular Collider-hadron hadron (FCC-hh)
\cite{FCC-hh1,FCC-hh2,FCC-hh3} could explore the structure of matter at an unprecedented level
of precision and are expected to provide answers to some of the most fundamental questions of basic science.
The FCC-hh also provides the basis for a lepton-hadron collider, such as the Future Circular Collider-lepton hadron
(FCC-eh) \cite{FCChe,Link-FCC-he-CERN}, which would be the cleanest, high-resolution microscope one can build
to resolve the substructure of matter. In addition to future hadron-hadron and lepton-hadron colliders, the future
lepton-lepton colliders are contemplated, such as the International Linear Collider (ILC) \cite{ILC-Brau}, the compact
linear collider (CLIC) \cite{CLIC-Burrows}, the circular electron-positron collider (CEPC) \cite{CEPC-Ahmad} and the
future circular collider $e^+e^-$ (FCC-ee) \cite{TLEP-Bicer} are considered. All these colliders contemplate in their
physics programs the study of the physics of the anomalous quartic gauge couplings (aQGC) and the anomalous triple gauge
couplings (aTGC) (which are deviations from the Standard Model (SM)), as well as the physics of the Higgs boson, top-quark,
tau-lepton and dark matter.

The measurements on the expected sensitivity to the aQGC $WWWW$, $WW\gamma\gamma$, $WW\gamma Z$, $WWZZ$, $ZZZZ$, $ZZZ\gamma$,
$ZZ\gamma\gamma$, $Z\gamma\gamma\gamma$, $\gamma\gamma\gamma\gamma$, as well as for the aTGC $WW\gamma$, $ZWW$, $HWW$, $HZZ$,
$HZ\gamma$, $H\gamma\gamma$ in hadron-hadron collisions provide an essential test of the electroweak sector of the SM. In the SM,
the non-Abelian nature of the $SU(2)_L \times U(1)_Y$ gauge group of the electroweak interactions leads to gauge boson self-interactions
through the QGC and the TGC.

On this subject, we study the signal process $pp \to W\gamma\gamma$, which includes the aQGC interactions via hadron-hadron scattering.
Our study has been performed assuming 1, 5, 10, 30 ${\rm ab^{-1}}$ of data collected at a center-of-mass energy of 100 TeV by the FCC-hh.
Additionally, we considered the systematic uncertainties of $\delta_{sys}=0\%, 3\%, 5\%, 10\%$ to estimate the sensitivity of the
$W\gamma\gamma$ signal, as well as on the aQGC. The process $pp \to W\gamma\gamma$, which is sensitive to the aQGC $WW\gamma\gamma$,
represents an evident starting point to look for the $W\gamma\gamma$ signal. Compared to those involving a higher number of heavy bosons,
this process requires a relatively low center-of-mass energy and gives rise to clean leptonic and hadronic final states suppressed by the
branching of only one massive vector boson $W$. Thus, the study of diphoton production in association with a massive gauge boson $W$ can
directly test the non-Abelian nature of the SM. Furthermore, the reaction $W\gamma\gamma$ was recently observed by the ATLAS Collaboration
at the LHC \cite{ATLAS-PRL2015,ATLAS-PRD2016}. This process probes the aQGC which offers a new and promising avenue of research into electroweak
symmetry breaking. The sensitivity to the aQGC $f_ {M,i}/\Lambda^4$ and $ f_ {T,j}/\Lambda^4$, (where $f$ represents the coupling of the
respective operator and $\Lambda$ represents the scale at which new physics appears), which are induced by dimension-8 operators,
are significantly better probed  by 2-4 orders of magnitude by the FCC-hh in comparison with the LHC, due to the stronger growth of the
anomalous cross-section with the center-of-mass energy.
These processes become rapidly more sensitive with increasing beam energy, providing strong motivation for a 100 TeV hadron-hadron
collider. It is appropriate to mention that the CMS Collaboration recently presented their measurements of the $pp \to  W^\pm\gamma\gamma$
and $pp \to Z\gamma\gamma$ cross-sections at 13 TeV and limits on anomalous quartic $WW\gamma\gamma$ couplings \cite{CMS-2105.12780}.
In another paper, the CMS Collaboration reported the observation of electroweak production of $W\gamma$ with two jets in proton-proton
collisions at $\sqrt{s}=13$ TeV and constraints are placed on the aQGC \cite{PLB811-2020}. The best experimental limits present reported
by the ATLAS and CMS Collaborations on the aQGC $W^+W^-\gamma\gamma$ at $\sqrt{s}=13$ TeV are summarized in Table I.

%Table 0
\begin{table}
\caption{Observed and expected at $95\%$ C.L. limits on the aQGC $W^+W^-\gamma\gamma$
reported by the ATLAS Collaboration at $\sqrt{s}=8$ TeV \cite{ATLAS-PRL2015} and the
CMS Collaborations at $\sqrt{s}=13$ TeV \cite{PLB811-2020,CMS-2105.12780}.}
\begin{tabular}{|c|c|c||c|c|}
\hline
\multicolumn{5}{|c|}{ATLAS Collaboration ($pp \to W(l\nu)\gamma\gamma + X$) \cite{ATLAS-PRL2015} } \\
\hline
%\multicolumn{5}{|c|}{ {$\sqrt{s}= 100$ TeV, \hspace{5mm} $\delta_{sys}=0\%$,  \hspace{5mm} $95\%$ C.L.} }\\
\hline
Couplings (TeV$^{-4}$)                & Observed                 &  Expected                           &      &   \\
\hline
n=0 \hspace{1mm}$f_{T0}/\Lambda^{4}$  & [-0.9; 0.9]$\times 10^2$ & [-1.2; 1.2]$\times 10^2$     &      &  \\
\hspace{8mm} $f_{M2}/\Lambda^{4}$  & [-0.8; 0.8] $\times 10^{4}$ & [-1.1; 1.1]$ \times 10^{4}$  &      &   \\
%\hline
\hspace{8mm} $f_{M3}/\Lambda^{4}$  & [-1.5; 1.4] $\times 10^{4}$ & [-1.9; 1.8]$ \times 10^{4}$  &      &  \\
\hline
n=1 \hspace{1mm}$f_{T0}/\Lambda^{4}$  & [-7.6; 7.3]$\times 10^2$ & [-9.6; 9.5]$\times 10^2$     &      &   \\
\hspace{8mm} $f_{M2}/\Lambda^{4}$  & [-4.4; 4.6] $\times 10^{4}$ & [-5.7; 5.9]$ \times 10^{4}$  &      &  \\
%\hline
\hspace{8mm} $f_{M3}/\Lambda^{4}$  & [-8.9; 8.0] $\times 10^{4}$ & [-11; 10]$ \times 10^{4}$    &      &   \\
\hline
n=2 \hspace{1mm}$f_{T0}/\Lambda^{4}$  & [-2.7; 2.6]$\times 10^3$ & [-3.5; 3.4]$\times 10^3$     &      & \\
\hspace{8mm} $f_{M2}/\Lambda^{4}$  & [-1.3; 1.3] $\times 10^{5}$ & [-1.6; 1.7]$ \times 10^{5}$  &      &   \\
%\hline
\hspace{8mm} $f_{M3}/\Lambda^{4}$  & [-2.9; 2.5] $\times 10^{5}$ & [-3.7; 3.3]$ \times 10^{5}$  &      &   \\
\hline\hline
\multicolumn{3}{|c||}{CMS Collaboration ($pp \to W\gamma jj$) \cite{PLB811-2020}} & \multicolumn{2}{|c|}{CMS Collaboration ($pp \to W^{\pm}\gamma\gamma$)    \cite{CMS-2105.12780}} \\
\hline
\hline
Couplings (TeV$^{-4}$)                & Observed               & Expected                & Observed         & Expected \\
\cline{1-5}
$f_{M0}/\Lambda^{4}$  & [-8.1; 8.0] & [-7.7; 7.6] & -  & - \\
\hline
$f_{M1}/\Lambda^{4}$  & [-12; 12]   & [-11; 11]   & -  & -  \\
\hline
$f_{M2}/\Lambda^{4}$  & [-2.8; 2.8] & [-2.7; 2.7] & [-39.9; 39.5] & [-57.3; 57.1]  \\
\hline
$f_{M3}/\Lambda^{4}$  & [-4.4; 4.4] & [-4.0; 4.1] & [-63.8; 65.0] & [-91.8; 92.6]  \\
\hline
$f_{M4}/\Lambda^{4}$  & [-5.0; 5.0] & [-4.7; 4.7] & - & -  \\
\hline
$f_{M5}/\Lambda^{4}$  & [-8.3; 8.3] & [-7.9; 7.7] & - & -  \\
\hline
$f_{M6}/\Lambda^{4}$  & [-16; 16]   & [-715; 15]  & - & -  \\
\hline
$f_{M7}/\Lambda^{4}$  & [-21; 20]   & [-19; 19]   & - & - \\
\hline\hline
$f_{T0}/\Lambda^{4}$  & [-0.6; 0.6] & [-0.6; 0.6] & [-1.30; 1.30] & [-1.86; 1.86]  \\
\hline
$f_{T1}/\Lambda^{4}$  & [-0.4; 0.4] & [-0.3; 0.4] & [-1.70; 1.66] & [-2.38; 2.38]  \\
\hline
$f_{T2}/\Lambda^{4}$  & [-1.0; 1.2] & [-1.0; 1.2] & [-3.64; 3.64] & [-5.16; 5.16]  \\
\hline
$f_{T5}/\Lambda^{4}$  & [-0.5; 0.5] & [-0.4; 0.4] & [-0.52; 0.60] & [-0.76; 0.84]  \\
\hline
$f_{T6}/\Lambda^{4}$  & [-0.4; 0.4] & [-0.3; 0.4] & [-0.60; 0.68] & [-0.92; 1.00]  \\
\hline
$f_{T7}/\Lambda^{4}$  & [-0.9; 0.9] & [-0.8; 0.9] & [-1.16; 1.16] & [-1.64; 1.72]  \\
\hline
\end{tabular}
\end{table}

Previous studies have been made of hadron-hadron, hadron-lepton, and lepton-lepton colliders by various authors and in different contexts.
For a review on the bounds on the aQGC in previous, present and future colliders, such as the LEP at the CERN, D0 and CDF at the Tevatron,
ATLAS and CMS at the LHC and in the post-LHC era as the LHeC and the FCC-he, the ILC, the CLIC, the CEPC and the FCC-ee, see references
\cite{ATLAS-PRL2015,ATLAS-PRD2016,ALEPH-Barate,DELPHI-Abreu,L3-Acciarri,OPAL-Abbiendi,CDF-Gounder,D0-Abbott,CMS-Chatrchyan,ATLAS-Aaboud,
LHeC-FCC-he-WWgg-Ari1,LHeC-FCC-he-WWgg-Ari2,Stirling,Leil,Bervan,Chong,Koksal,Stirling1,Atag,Eboli1,Sahin,Koksal1,Chapon,Koksal2,
Senol,Koksal3,Yang,Eboli2,Eboli4,Bell,Ahmadov,Schonherr,Wen,Ye,Perez,Sahin1,Senol1,Baldenegro,Fichet,Pierzchala,Gutierrez,Belanger,Aaboud,
Eboli,Eboli3,Gutierrez-EPJC81-2021,twiki.cern,Eboli-PRD101-2020}.

The paper is organized as follows. In Section II, we briefly present the theory related to the $pp \to W\gamma\gamma$ process
and the approach to deal with the anomalous couplings. The study of the $W\gamma\gamma$ signal, as well as the future projections
on the $f_ {M,i}/\Lambda^4$ and $ f_ {T,j}/\Lambda^4$ couplings at $95\%$ C.L. in hadron-hadron interactions at the FCC-hh are discussed
in Section III. Finally, we summarize our conclusions in Section IV.

\section{Model-independent general formalism for the aQGC for the $pp \to W\gamma \gamma $ signal}

Effective Field Theories (EFT) are a powerful tool in quantum field theory and share many advantages of both fundamental
theories and phenomenological models. The theory has the same field content and the same linearly-realized $SU(3)_C \times SU(2)_L
\times U(1)_Y$ local symmetry as the SM. Besides, cover a wide range of phenomena in particle physics and can be translated
into an effective Lagrangian approach for the study of the aTGC/aQGC.

The deviations from the SM can be analyzed in the context of the direct searches for new physics or in a  model-independent
approach for the test of Electroweak Symmetric Breaking (EWSB) through the effective Lagrangian approach. In this regard,
low energy effects from new  physics at scales beyond the current kinematic reach of the LHC can be parameterized by an EW effective
Lagrangian.

The formalism of the aQGC has been widely discussed in the literature \cite{Eboli1,Eboli2,Eboli4,Eboli,Degrande,Eboli3,Eboli-PRD101-2020}.
The effective Lagrangian that gives all quartic couplings is given as follows

\begin{equation}
{\cal L}_{eff}= {\cal L}_{SM} +\sum_{k=0}^1\frac{f_{S, k}}{\Lambda^4}O_{S, k} + \sum_{i=0}^{7}\frac{f_{M, i}}{\Lambda^4}O_{M, i}
+\sum_{j=0,1,2,5,6,7,8,9}^{}\frac{f_{T, j}}{\Lambda^4}O_{T, j},
\end{equation}

\noindent where each $O_{S, k}$, $O_{M, i}$ and $O_{T, j}$ is a gauge-invariant operator of dimension-8 and $\frac{f_{S, k}}{\Lambda^4}$,
$\frac{f_{M, i}}{\Lambda^4}$ and $\frac{f_{T, j}}{\Lambda^4}$ are the corresponding effective coefficient. According to the structure
of the operator, there are three classes of genuine aQGC operators  as shown in Eq. (1) \cite{Degrande, Eboli3}.  \\

$\bullet$ The first set of operators $O_{S, k}$ contains only $D_\mu\Phi$.

These dimension-eight operators induce the following quartic $WWWW$, $WWZZ$, and $ZZZZ$ interactions and explicitly are given by

\begin{eqnarray}
O_{S, 0}=    & [(D_\mu\Phi)^\dagger (D_\nu\Phi)]\times [(D^\mu\Phi)^\dagger (D^\nu\Phi)],  \nonumber \\
O_{S, 1}=    &  [(D_\mu\Phi)^\dagger (D^\mu\Phi)]\times [(D_\nu\Phi)^\dagger (D^\nu\Phi)].
\end{eqnarray}

$\bullet$ The second set of operators $O_{M, i}$ contains $D_\mu\Phi$ and the field strength tensors $W^{\mu\nu}$ and $B^{\mu\nu}$.

In this type of dimension-eight operators, are induce the quartic $WWWW$, $WWZZ$, $ZZZZ$, $WW\gamma Z$, $WW\gamma\gamma$,
$ZZZ\gamma$ and $ZZ\gamma\gamma$ interactions. The explicit form of these operators is as follows:

\begin{eqnarray}
O_{M, 0}&=   & Tr[W_{\mu\nu} W^{\mu\nu}]\times [(D_\beta\Phi)^\dagger (D^\beta\Phi)], \nonumber   \\
O_{M, 1}&=   & Tr[W_{\mu\nu} W^{\nu\beta}]\times [(D_\beta\Phi)^\dagger (D^\mu\Phi)],  \nonumber \\
O_{M, 2}&=   & [B_{\mu\nu} B^{\mu\nu}]\times [(D_\beta\Phi)^\dagger (D^\beta\Phi)], \nonumber \\
O_{M, 3}&=   & [B_{\mu\nu} B^{\nu\beta}]\times [(D_\beta\Phi)^\dagger (D^\mu\Phi)], \nonumber  \\
O_{M, 4}&=   & [(D_\mu\Phi)^\dagger W_{\beta\nu} (D^\mu\Phi)]\times B^{\beta\nu},  \\
O_{M, 5}&=   & [(D_\mu\Phi)^\dagger W_{\beta\nu} (D^\nu\Phi)]\times B^{\beta\mu},  \nonumber \\
O_{M, 6}&=   & [(D_\mu\Phi)^\dagger W_{\beta\nu} W^{\beta\nu} (D^\mu\Phi)],  \nonumber \\
O_{M, 7}&=   & [(D_\mu\Phi)^\dagger W_{\beta\nu} W^{\beta\mu} (D^\nu\Phi)].  \nonumber
\end{eqnarray}

$\bullet$ The third set of operators $O_{T, j}$ contains only the field strength tensors $W^{\mu\nu}$ and $B^{\mu\nu}$.

For this class of dimension-eight operators, are induce the quartic $WWWW$, $WWZZ$, $ZZZZ$, $WW\gamma Z$, $WW\gamma\gamma$,
$ZZZ\gamma$, $ZZ\gamma\gamma$,  $ZZ\gamma\gamma$, and  $\gamma\gamma\gamma\gamma$ interactions. These operators are explicitly
given by:

\begin{eqnarray}
O_{T, 0} &=  &  Tr[W_{\mu\nu} W^{\mu\nu}]\times Tr[W_{\alpha\beta}W^{\alpha\beta}],  \nonumber \\
O_{T, 1} &=  &  Tr[W_{\alpha\nu} W^{\mu\beta}]\times Tr[W_{\mu\beta}W^{\alpha\nu}],  \nonumber \\
O_{T, 2} &=  &  Tr[W_{\alpha\mu} W^{\mu\beta}]\times Tr[W_{\beta\nu}W^{\nu\alpha}],  \nonumber \\
O_{T, 5} &=  &  Tr[W_{\mu\nu} W^{\mu\nu}]\times B_{\alpha\beta}B^{\alpha\beta},  \nonumber \\
O_{T, 6} &=  &  Tr[W_{\alpha\nu} W^{\mu\beta}]\times B_{\mu\beta}B^{\alpha\nu},  \\
O_{T, 7} &=  &  Tr[W_{\alpha\mu} W^{\mu\beta}]\times B_{\beta\nu}B^{\nu\alpha},  \nonumber \\
O_{T, 8} &=  &  B_{\mu\nu} B^{\mu\nu}B_{\alpha\beta}B^{\alpha\beta},  \nonumber \\
O_{T, 9} &=  &  B_{\alpha\mu} B^{\mu\beta}B_{\beta\nu}B^{\nu\alpha}.   \nonumber
\end{eqnarray}

In Eqs. (2)-(4), $\Phi$ stands for the Higgs doublet, and the covariant derivative of the Higgs field is given by
$D_\mu\Phi=(\partial_\mu + igW^j_\mu \frac{\sigma^j}{2} + \frac{i}{2}g'B_\mu )\Phi$, and $\sigma^j (j=1,2,3)$ represent
the Pauli matrices, while $W^{\mu\nu}$ and $B^{\mu\nu}$ are the gauge field strength tensors for $SU(2)_L$ and $U(1)_Y$,
respectively. It is appropriate to mention that the limits of LEP2 on the $WW\gamma\gamma$ vertex \cite{Data2020} described
in terms of the anomalous $a_0/\Lambda^2$ and $a_c/\Lambda^2$ couplings can be translated into limits on $f_{M,i}/\Lambda^4$
with $i = 0-7$. The $\frac{f_{M, i}}{\Lambda^4}$ Wilson coefficients are related with $a^W_{0,c}/\Lambda^2$ couplings as follows
\cite{Belanger,Eboli3,Degrande,Baak}:

\begin{eqnarray}
\frac{f_{M, 0}}{\Lambda^4}&=&\frac{a_0}{\Lambda^2}\frac{1}{g^2v^2},  \\
\frac{f_{M, 1}}{\Lambda^4}&=&-\frac{a_c}{\Lambda^2}\frac{1}{g^2v^2},  \\
\frac{f_{M, 0}}{\Lambda^4}&=&\frac{f_{M, 2}}{2\Lambda^4}= \frac{f_{M, 4}}{\Lambda^4}=\frac{f_{M, 6}}{2\Lambda^4},    \\
\frac{f_{M, 1}}{\Lambda^4}&=&\frac{f_{M, 3}}{2\Lambda^4}= -\frac{f_{M, 5}}{2\Lambda^4}=-\frac{f_{M, 7}}{2\Lambda^4}.
\end{eqnarray}

It is important to note that the operators in Eq. (1) lead to unitarity violation. This is from the truncation of the SM-EFT
series at any finite order, which corresponds to the energy region above which the contribution of the anomalous couplings are
largely suppressed. The EFT is not a complete model and violates unitarity at sufficiently high energy scales. The form factor 
scheme is a scheme where the effects ultimately depend on the value of the cut-off scale. The standard procedure to avoid
this unphysical behavior of the cross-section and to obtain meaningful limits is to multiply the anomalous couplings by a dipole
form factor of the form \cite{Ellison-ARNPS1998,Eboli4}:

\begin{equation}
FF=\frac{1}{(1+\frac{\hat s}{\Lambda^2_{FF}})^2}.
\end{equation}

\noindent In Eq. (9), $\hat s$ corresponds to the squared invariant mass of the produced bosons, and $\Lambda_{FF}$
is the cut-off scale of the dipole form factor, which corresponds to the energy regime above which the contributions
of the anomalous couplings are largely suppressed. Nevertheless, because of selection of form factor is arbitrary
in the literature, we will examine aQGC parameters without using any form factor in this study.

\section{Projections on the cross-sections of the $pp \to W\gamma \gamma$ signal at the FCC-hh}

We study the process of simple $W$ production in association with two-photon observed by the ATLAS Collaboration at the LHC
\cite{ATLAS-PRL2015,ATLAS-PRD2016}. The schematic diagram is given in Fig. 1. The double production of photons may take place
due to the process

\begin{equation}
p p \to  W\gamma\gamma.
\end{equation}

The Feynman diagrams of the subprocess:

\begin{equation}
q\bar q \to W\gamma\gamma,
\end{equation}

\noindent are shown in Fig. 2. In Eq. (11), $q \bar q = u \bar u, d \bar d, c \bar c, s \bar s, b \bar b$,
$W^\pm \to l\nu_l$, $l\nu_l = e^-\nu_e, \mu\nu_\mu$ (for leptonic decay) and $W^\pm \to jj$ (for hadronic decay),
respectively. The subprocess $q \bar q \to W\gamma \gamma$ is described by 6 tree-level Feynman diagrams,
where diagram (1) exhibit the dependence on the quartic ($WW\gamma\gamma$) gauge boson couplings.

Our calculations on the total cross-sections at leading-order (LO) for the process $pp \to W\gamma\gamma$ are realized
with the help of the MadGraph5\_aMC@NLO \cite{MadGraph} package, where the operators given in Eqs. (3) and (4) are implemented
into MadGraph5\_aMC@NLO through Feynrules package \cite{AAlloul} as a Universal FeynRules Output (UFO) module \cite{CDegrande}.
Furthermore, we consider the NN23Lo1 parton distribution functions (PDF) \cite{PDF-JHEP07-2002}. We apply the following kinematic
cuts on the rapidity, $\eta^{\gamma, l, j}$, and the transverse momenta $p^{l, \gamma, j}_T$, of the charged leptons, hadrons,
and the photon, as well as on the distance between particles $\Delta R^{\gamma\gamma,\gamma l,jj,\gamma j}$ in our computation
to obtain high signal efficiency together with good background rejection \cite{ATLAS-PRL2015,PLB811-2020,CMS-2105.12780}.   \\

$\bullet$ Selected cuts for the leptonic decay of the $W$-boson:

\begin{eqnarray}
%\begin{array}{c}
p^l_T &>& 40\hspace{0.8mm}{\rm GeV} \hspace{3mm} \mbox{(minimum $p_T$ for the leptons)},   \nonumber \\
p^\gamma_T &>& 250\hspace{0.8mm} {\rm GeV} \hspace{3mm} \mbox{(minimum $p_T$ for the photons)},    \nonumber \\
|\eta^{\gamma}| &<& 2.5 \hspace{3mm} \mbox{(maximum rapidity for the photons)},             \\
|\eta^{l}| &<& 2.5 \hspace{3mm} \mbox{(maximum rapidity for the leptons)},     \nonumber        \\
\Delta R^{\gamma\gamma}_{\rm min} &=& 1.5 \hspace{3mm} \mbox{(minimum distance between photons)},  \nonumber \\
\Delta R^{\gamma\gamma}_{\rm max} &=& 6.0 \hspace{3mm} \mbox{(maximum distance between photons)},   \nonumber  \\
\Delta R^{\gamma l}_{\rm min} &=& 0.4 \hspace{3mm} \mbox{(minimum distance between lepton and photon)}.   \nonumber
%\end{array}
\end{eqnarray}

$\bullet$ Selected cuts for the hadronic decay of the $W$-boson:

\begin{eqnarray}
%\begin{array}{c}
p^j_T &>& 40\hspace{0.8mm}{\rm GeV} \hspace{3mm} \mbox{(minimum $p_T$ for the jets)},  \nonumber  \\
p^\gamma_T &>& 250\hspace{0.8mm} {\rm GeV} \hspace{3mm} \mbox{(minimum $p_T$ for the photons)},   \nonumber  \\
|\eta^j| &<& 5 \hspace{3mm} \mbox{(maximum rapidity for the jets)},     \nonumber  \\
|\eta^{\gamma}| &<& 2.5 \hspace{3mm} \mbox{(maximum rapidity for the photons)},  \nonumber  \\
60\hspace{0.8mm} {\rm GeV} < M_{jj} &<& 100 \hspace{0.8mm}{\rm GeV} \hspace{3mm} \mbox{(invariant mass of two jets)},   \\
\Delta R^{jj}_{\rm min} &=& 0.5 \hspace{3mm} \mbox{(minimum distance between jets)},   \nonumber  \\
\Delta R^{jj}_{\rm max} &=& 3.5 \hspace{3mm} \mbox{(maximum distance between jets)},    \nonumber    \\
\Delta R^{\gamma\gamma}_{\rm min} &=& 1.5 \hspace{3mm} \mbox{(minimum distance between photons)},     \nonumber \\
\Delta R^{\gamma\gamma}_{\rm max} &=& 6.0 \hspace{3mm} \mbox{(maximum distance between photons)},     \nonumber   \\
\Delta R^{\gamma j}_{\rm max} &=& 0.4 \hspace{3mm} \mbox{(minimum distance between photons and jets)}.        \nonumber
%\end{array}
\end{eqnarray}

Here, the SM process with the final state should be accepted as a background for the process $pp \to W\gamma\gamma$
with leptonic and hadronic decay channels. Additionally, we have considered the following background processes $\gamma\gamma jj$,
$\gamma\gamma \gamma jj$ and $\gamma\gamma jjj$ for the hadronic decay channel of the $W$-boson.

The $W\gamma\gamma$ final state is characterized by the presence of the $W$ decay products and two photons. For our analysis,
the muon, electron, and hadron $W$ decays are treated as the signal. In this regard, it is appropriate to mention that the $W \to jj'$
decay has a very low signal-to-noise ratio since it is easily faked by the di-jet production.

On the other hand, the measurement of the cross-section and the anomalous $WW\gamma\gamma$ couplings are affected by
a series of systematic uncertainties. The systematic uncertainties are connected to the method that has been implemented to estimate
a particular contribution. Systematic uncertainties in the measured cross-sections arise from uncertainties in the physics object
reconstruction and identification, the procedures used to correct for detector effects, the background estimation, the usage of
theoretical cross-sections for signal and background processes, the PDF, and luminosity. In this regard, the systematic
uncertainties should be taken into account and discussed in order to quantify the impact on the total cross-section,
as well as in the anomalous $WW\gamma\gamma$ couplings in order to make more realistic predictions.

In Ref. \cite{ATLAS-PRL2015}, the ATLAS Collaboration report evidence on the production $pp \to W\gamma\gamma +X$, which is
accessible for the first time with 8 TeV LHC data set. In this paper, the measured cross-section is used to set limits on
aQGC $WW\gamma\gamma$. Its bounds on the aQGC $WW\gamma\gamma$ are obtained from a selection of the most relevant systematic
uncertainties, which are of the order of $1-8\%$ (see Tables I and III of this reference).

Measurements of the total $Z\gamma jj$ production cross-section in proton-proton collisions with the CMS detector and limits
on anomalous triple-gauge-boson couplings are obtained in Ref. \cite{CMS-JHEP06-2020}. The systematic uncertainties from
the reconstruction of the events in the detector and the background subtraction are summarized in Table 2 of the extracted
signal. The main sources of systemic uncertainties are the systematic uncertainties in the trigger, lepton reconstruction,
and selection efficiencies are measured using the tag-and-probe technique and are $2-3\%$. An uncertainty of $2.5\%$ in the
integrated luminosity. The PDF uncertainty in the QCD-induced $Z\gamma jj$ events is in the range of $1-3\%$. All the
above systematic uncertainties are applied to both the measured significance of the signal and the search for aQGC.

Based on these results, and in order to obtain the sensitivity on the anomalous $WW\gamma\gamma$ couplings we choose
the systematic uncertainties of $\delta_{sys}= 0\%, 3\%, 5\%$ and $10\%$ in correspondence with the systematic
uncertainties assume in Refs. \cite{ATLAS-PRL2015,CMS-JHEP06-2020,PLB811-2020,HL-LHC-HE-LHC,FCC-hh2,EPJC80-2020,
CMS-PAS-SMP-20-016,CMS-PAS-SMP-19-013}.

On the other hand, it is known that the high dimensional operators could affect the $p^{\gamma}_T$ photon transverse momentum,
especially in the region with a large $p^{\gamma}_T$ values, which can be very useful to distinguish signal and background events
(see Subsection IIIA).

We have carried out a study of the total cross-section $\sigma_{Tot}=\sigma_{Tot}(\sqrt{s}, \frac{f_{M, i}}{\Lambda^4},
\frac{f_{T, j}}{\Lambda^4})$ for the double photon production channel in association with the $W$ boson $p \bar p \to W\gamma \gamma$.
It should be mentioned that the fiducial phase space for this measurement is defined by the requirements of photons transverse momentum
$p^\gamma_T$, leptons transverse momentum $p^l_T$, jets transverse momentum $p^j_T$, photons pseudorapidity $\eta^\gamma$ and jets
pseudorapidity $\eta^j$. The pseudorapidity requirement reduces the contamination from other particles misidentified as photons or jets.

The variation of the production cross-sections of the $W\gamma\gamma$ signal as a function of the center-of-mass energy $\sqrt{s}$
of the FCC-hh and taking one anomalous coupling at a time that is $f_{M,i}/\Lambda^4$ or $f_{T,j}/\Lambda^4$ are shown in Figs. 3-4.
The curves depict the cross-section for $pp \to  W\gamma\gamma$ for the leptonic and hadronic decay channel of the $W$-boson. These
figures show a stronger cross-section dependence with respect to the center-of-mass energy of the FCC-hh, as well as with the anomalous
$f_{M,i}/\Lambda^4$ and $f_{T,j}/\Lambda^4$ couplings in the range allowed by these parameters. Our results indicate that the cross-sections
as a function of $\sqrt{s}$ are higher for the proposed 100 TeV energy and detector acceptance cuts (see Eqs. (12)-(13)).

The total cross-sections of the $p p \to  W\gamma\gamma$ signal are presented as a function of $\frac{f_{M, i}}{\Lambda^4}$
and $\frac{f_{T, j}}{\Lambda^4}$ in Figs. 5 and 6 for the center-of-mass energy of $\sqrt{s}= 100$ TeV. The production
is identified through the leptonic and hadronic decay modes of the $W$-boson, respectively. These total cross-sections
$\sigma(p p \to  W\gamma\gamma ) =\sigma(\sqrt{s}, \frac{f_{M, i}}{\Lambda^4}, \frac{f_{T, j}}{\Lambda^4})$ clearly show
a strong dependence on the anomalous $\frac{f_{M, i}}{\Lambda^4}$ and $\frac{f_{T, j}}{\Lambda^4}$ couplings as well as with
the center-of-mass energy of the FCC-hh. For instance,
the total cross-section for $\frac{f_{M, 1}}{\Lambda^4} = \frac{f_{T, 2}}{\Lambda^4}=\pm 5\times 10^{-10}$ GeV$^{-4}$ are the following:
$\sigma(\sqrt{s},\frac{f_{M, 1}}{\Lambda^4})= 2.16\times 10^{-1}$ pb and $\sigma(\sqrt{s},\frac{f_{T, 2}}{\Lambda^4})= 1.52\times 10^{5}$ pb for
$\sqrt{s}= 100$ TeV in the leptonic decay channel of the $W$-boson. While the total cross-sections for $\frac{f_{M, 1}}{\Lambda^4}
= \frac{f_{T, 2}}{\Lambda^4}=\pm 5\times 10^{-10}$ GeV$^{-4}$ in the hadronic decay channel of the $W$-boson are as follows:
$\sigma(\sqrt{s},\frac{f_{M, 1}}{\Lambda^4})= 2.64\times10^{-3}$ pb and $\sigma(\sqrt{s}, \frac{f_{T, 2}}{\Lambda^4})= 7.15\times10^{2}$ pb for $\sqrt{s}= 100$ TeV. By comparison, the total cross-section for the process $p p \to  W\gamma\gamma$ at the LHC with $\sqrt{s}=14$ TeV is of $\sigma(p p \to  W\gamma\gamma)
= 7.25$ fb \cite{PRD83-2011}, which is of the same order of magnitude with respect to the prediction of the FCC-hh for the hadronic decay channel of the $W$-boson. Figs. 5 and 6 show that the effect of the $\frac{f_{T, j}}{\Lambda^4}$ couplings on the total cross-sections is stronger than that
corresponding to the $\frac{f_{M, i}}{\Lambda^4}$ couplings, as clearly shown by the solid-lines and dashed-lines. The difference comes from the fact that $O_{T}$ and $O_{M}$ have different momentum dependence.

The effect on the total cross-sections of the $\frac{f_{T, j}}{\Lambda^4}$ coupling with respect to the $\frac{f_{M, i}}{\Lambda^4}$
coupling is of order of magnitude ${\cal O}(10^6)$. These results illustrate that the total cross-section increases with the increase in the
center-of-mass energy of the collider. The deviation from the SM of the total cross-section containing anomalous couplings at $\sqrt{s}= 100$ TeV
is larger than at $\sqrt{s} = 14$ TeV, both including $\frac{f_{M, i}}{\Lambda^4}$ and $\frac{f_{T, j}}{\Lambda^4}$. Thus, the obtained limits
on the anomalous $\frac{f_{M, i}} {\Lambda^4}$ and $\frac{f_{T, j}}{\Lambda^4}$ couplings at $\sqrt{s}= 100$ TeV are expected
to be more restrictive than the limits at $\sqrt{s} = 14$ TeV.

\subsection{Future projections on the anomalous $\frac{f_{M, i}}{\Lambda^4}$ and $\frac{f_{T, j}}{\Lambda^4}$ couplings at the FCC-hh}

As we aforementioned, the processes involving two photons and one $W$ in the final state, namely, the production
of $pp \to W\gamma\gamma$ are relevant for studying anomalous gauge interactions, as they give direct access to
quartic $WW\gamma\gamma$ couplings. Also, the diphoton in the final state of the $pp \to W\gamma\gamma$ signal
at the FCC-hh has the advantage of being identifiable with high purity and efficiency. Thus, the diphoton channels
are especially sensitive for new physics BSM regarding modest backgrounds and excellent mass resolution.

It is appropriate to mention that in Figs. 7 and 8, we take into account the present LHC and CMS limits on $f_{M,i}/\Lambda^4$
and $f_{T,j}/\Lambda^4$ Wilson coefficients \cite{ATLAS-PRL2015,CMS-2105.12780,PLB811-2020,twiki.cern}. Our best sensitivity
on $f_{M,i}/\Lambda^4$ and $f_{T,j}/\Lambda^4$ couplings as given in Table II are on the orders of $10^{-1}$ and $10^{-4}$,
that is $f_{M,2}/\Lambda^4 =[-2.21; 2.42]\times 10^{-1} \hspace{1mm}{\rm TeV}^{-4}$ and $f_{T,5}/\Lambda^4 =[-3.09; 3.34]
\times 10^{-4}\hspace{1mm}{\rm TeV}^{-4}$ for the center-of-mass energy of 100 TeV and integrated luminosity of $30\hspace{1mm}
{\rm ab^{-1}}$. Therefore, since our limits on $f_{M,i}/\Lambda^4$ couplings are obtained worse that in $f_{T,j}/\Lambda^4$
couplings, we have taken as $f_{M,i}/\Lambda^4 = 100\hspace{1mm}{\rm TeV}^{-4}= 1\times 10^{-10}\hspace{1mm}{\rm GeV}^{-4}$
and $f_{T,j}/\Lambda^4 = 1\hspace{1mm}{\rm TeV}^{-4} = 1\times 10^{-12} \hspace{1mm}{\rm GeV}^{-4}$ to obtain Fig. 7.
While to obtain Fig. 8, we have taken $f_{M,i}/\Lambda^4 = 1\times 10^{-9}\hspace{1mm}{\rm GeV}^{-4}$
and $f_{T,j}/\Lambda^4 = 1\times 10^{-9} \hspace{1mm}{\rm GeV}^{-4}$.

The solid lines in Fig. 7 show the number of expected events as a function of the $p^\gamma_T$ photon transverse momentum
for the $p p \to W\gamma \gamma$ signal, while the dashed lines in this figure represent the number of expected events as a function
of the $p^\gamma_T$ photon transverse momentum for the $W\gamma\gamma$ background at the FCC-hh with $\sqrt{s} = 100$ TeV.
The distributions are for $f_{M,i}/\Lambda^4$
with $i= 0, 1, 2, 3, 4, 5, 7$, $f_{T,j}/\Lambda^4$ with $j=0, 1, 2, 5, 6, 7$ and backgrounds for the leptonic decay channel of the
$W$-boson. Fig. 8 shows the number of expected events from the signal and background Monte Carlo simulations for an integrated
luminosity corresponding to 100 ${\rm fb}^{-1}$. These results are for the hadronic decay channel of the $W$-boson. Fig. 8 also
shows the background results separately, that is, $\gamma\gamma jj$, $\gamma\gamma\gamma jj$, and $\gamma\gamma jjj$.
As can be seen from Figs. 7 and 8, when the photon in the final state is applied as $p^\gamma_T > 250$ GeV, backgrounds
are significantly suppressed, and the signal becomes more dominant.

As aforementioned, Figs. 7 and 8 show the number of expected events as a function of the $p_T(\gamma)$ transverse momentum for the
$p p \to W \gamma \gamma$ signal and backgrounds at the FCC-hh. The distribution for the signal clearly shows great sensitivity with
respect to the anomalous $f_{M,2}/\Lambda^4$ and $f_{T,5}/\Lambda^4$ couplings for both cases leptonic and hadronic for
$p_T(\gamma) > 250$ GeV, which is more pronounced at the high-energy tails of some distributions. In addition, we must take into account
the fact that the operators ${\cal O}_M$ and ${\cal O}_T$ have different momentum dependence.

In summary, Figs. 7 and 8 show the reconstructed $p^\gamma_T$ photon transverse momentum distributions of the $p p \to W\gamma \gamma$
signal and the total background for $100$ ${\rm}fb^{-1}$ with $\sqrt{s}=100$ TeV for the leptonic decay channel and the hadronic decay
channel of the $W$-boson, respectively. The distributions given by Figs. 7 and 8 clearly show great sensitivity concerning the anomalous
$f_{M,i}/\Lambda^4$ and $f_{T,j}/\Lambda^4$ couplings for both cases leptonic and hadronic. The analysis of these distributions is important
to discriminate the basic acceptance cuts for $W\gamma\gamma$ events at the FCC-hh.

The projections on the anomalous $\frac{f_{M, i}}{\Lambda^4}$ and $\frac{f_{T, j}}{\Lambda^4}$ couplings at $95\%$ C. L.
with $\sqrt{s}=100$ TeV, ${\cal L}=1, 5, 10, 30$ ${\rm ab^{-1}}$ and systematic uncertainties of $\delta_{sys} = 0\%, 3\%, 5\%, 10\%$
\cite{ATLAS-PRL2015,CMS-JHEP06-2020,PLB811-2020,HL-LHC-HE-LHC,FCC-hh2,EPJC80-2020,CMS-PAS-SMP-20-016,CMS-PAS-SMP-19-013}
are extracted using the methodology of the $\chi^2$ function \cite{Koksal6,Gutierrez2,Billur5,Koksal7,Gutierrez3}:

\begin{equation}
\chi^2(f_{M,i}/\Lambda^4, f_{T,j}/\Lambda^4)=\Biggl(\frac{\sigma_{SM}(\sqrt{s})-\sigma_{BSM}(\sqrt{s}, f_{M,i}/\Lambda^4, f_{T,j}/\Lambda^4)}
{\sigma_{SM}(\sqrt{s})\sqrt{(\delta_{st})^2 + (\delta_{sys})^2}}\Biggr)^2,
\end{equation}

\noindent where $\sigma_{SM}(\sqrt{s})$ is the cross-section in the SM and $\sigma_{BSM}(\sqrt{s}, f_{M,i}/\Lambda^4, f_{T,j}/\Lambda^4)$
is the cross-section in the presence of BSM interactions, and $\delta_{st}=\frac{1}{\sqrt{N_{SM}}}$ is the statistical
error and $\delta_{sys}$ is the systematic error. The number of events is given by $N_{SM}={\cal L}_{int}\times \sigma_{SM}$,
where ${\cal L}_{int}$ is the integrated luminosity of the FCC-hh. When performing the analysis for the hadronic final state, we take into account the processes $pp\to\gamma\gamma jj$ (SM), $\gamma\gamma\gamma jj$, $\gamma\gamma jjj$ as seen in Fig. 8. With the cuts given in Eq. (13), we obtain the cross sections for the processes $pp\to\gamma\gamma jj$ (SM), $\gamma\gamma\gamma jj$, $\gamma\gamma jjj$ as $1.15\times10^{-2}$, $3.81\times10^{-6}$ and $3.07\times10^{-5}$ pb, respectively. The SM background is three orders of magnitude and four orders of magnitude larger than the processes $pp\to\gamma\gamma jjj$ and $pp\to\gamma\gamma\gamma jj$. Because of this, while obtaining the sensitivities of the anomalous $f_{M,i}/\Lambda^4$ and $f_{T,j}/\Lambda^4$ couplings, the effects of backgrounds the processes $pp\to\gamma\gamma jjj$ and $pp\to \gamma\gamma\gamma jj$ did not take into account. Therefore we consider only total cross-section with SM for hadronic final state.

It is appropriate to mention that we examine the sensitivity limits on the anomalous $f_{M,i}/\Lambda^4$ and $f_{T,j}/\Lambda^4$
couplings by considering each anomalous coupling separately. In addition, we have used the cumulative distribution function for the
$\chi^2$, which has been defined for use with a least-squares method. We have taken into account the values from Table 40.2 in \cite{Data2020}
for coverage probability in the large data sample limit for one parameter. As can be seen from that table, the limit obtained
at the $95\%$ C.L. on an one-dimensional aQGC parameter is performed by the $\chi^2$ test, which is equal to 3.84.

In Tables II-III, we present the expected projections for the anomalous $\frac{f_{M, i}}{\Lambda^4}$ and $\frac{f_{T, j}}{\Lambda^4}$
couplings at $95\%$ C.L. for the run of the FCC-hh with $\sqrt{s}= 100$ TeV and ${\cal L}=1, 5, 10, 30$ ${\rm ab^{-1}}$.
In addition, we consider the systematic uncertainties given by the representative values of $\delta_{sys} = 0\%, 3\%, 5\%, 10\%$
\cite{ATLAS-PRL2015,CMS-JHEP06-2020,PLB811-2020,HL-LHC-HE-LHC,FCC-hh2,EPJC80-2020,CMS-PAS-SMP-20-016,CMS-PAS-SMP-19-013}. Here, any
couplings are calculated while fixing the other couplings to zero. From our results given in Tables II-III, we observed that our
sensitivity constraints on the anomalous $\frac{f_{M, i}}{\Lambda^4}$ and $\frac{f_{T, j}}{\Lambda^4}$ parameters are a factor
of the order of magnitude ${\cal O}(10^{-3})$ more sensitive than the limits obtained for the ATLAS and CMS Collaborations given
in Table I \cite{ATLAS-PRL2015,CMS-2105.12780,PLB811-2020}, as well as by others authors \cite{CMS-JHEP08,CMS-JHEP06-2017,ATLAS-PRD94-2016,
CMS-JHEP06-2020}. Our limits on the anomalous $\frac{f_{M, i}}{\Lambda^4}$ and $\frac{f_{T, j}}{\Lambda^4}$ couplings obtained through the
process $pp \to W\gamma \gamma$ for the leptonic decay channel of the $W$-boson in the final state and with the FCC-hh parameters, are the
most stringent to date. There are several aspects to why our limits obtained in our work on the anomalous $f_ {M,i}/\Lambda^4$ and
$ f_ {T,j}/\Lambda^4$ couplings are most stringent than those reported in the literature. One of these is the total cross-section of the
$pp \to W\gamma\gamma$ process at LO, and others are due to the high energy of the center-of-mass and high integrated luminosities
of the FCC-hh, that is, $\sqrt{s}=100$ TeV and ${\cal L}=1, 5, 10, 30$ ${\rm ab}^{-1}$.

A similar study is done on the expected projections for the $\frac{f_{M, i}}{\Lambda^4}$ and $\frac{f_{T, j}}{\Lambda^4}$ couplings
at the FCC-hh with $\sqrt{s}= 100$ TeV, ${\cal L}=1, 5, 10, 30$ ${\rm ab^{-1}}$ and $\delta_{sys} = 0\%, 3\%, 5\%, 10\%$ at $95\%$ C.L.
for the hadronic decay channel of the $W$-boson in the final state, as is shown in Tables IV-V.

Tables II-V show that the limits with increasing the luminosity on the anomalous couplings do not increase proportionately
to the luminosity due to the systematic error considered here. The reason for this situation is the systematic error which
is much bigger than the statistical error. If the systematic error is reduced, we expect better limits on the couplings.
For instance, in the case of leptonic decay, our best limits on the anomalous couplings for the process $pp \to W\gamma\gamma$
with $\sqrt{s} = 100$ TeV, ${\cal L} = 30$ ${\rm ab^{-1}}$ and $\delta_{sys} = 0\%$ can be almost improved up to 3 times for
$f_{T,j}/\Lambda^4$  and $f_{M,i}/\Lambda^4$ according to case with $\delta_{sys} = 10\%$. Similarly, in the case of hadronic
decay, the best limits on $f_{T,j}/\Lambda^4$  and  $f_{M,i}/\Lambda^4$ with $\delta_{sys} = 0\%$ can set more stringent up to
4 times better than the limits obtained with $\delta_{sys} = 10\%$.

%Table 1
\begin{table}[h]
\caption{Sensitivity at $95\%$ C.L. on the aQGC  $W^+W^-\gamma\gamma$ of the $pp \to W\gamma\gamma$ signal for $\sqrt{s}=100$ TeV, ${\cal L}=1, 5, 10,30$ ${\rm ab^{-1}}$ and $\delta_{sys}= 0\%, 3\%$ at the FCC-hh.}
\begin{tabular}{|c|c|c|c|c|c|}
\hline
\multicolumn{5}{|c|}{Leptonic channel} \\
\hline
\multicolumn{5}{|c|}{ {$\sqrt{s}= 100$ TeV, \hspace{5mm} $\delta_{sys}=0\%$,  \hspace{5mm} $95\%$ C.L.} }\\
\hline
Couplings (TeV$^{-4}$) & 1 ab$^{-1}$ & 5 ab$^{-1}$ & 10 ab$^{-1}$ & 30 ab$^{-1}$ \\
\hline
$f_{M0}/\Lambda^{4}$  & [-3.77; 3.32] & [-2.60; 2.15] & [-2.23; 1.77] & [-1.75; 1.30] \\
\hline
$f_{M1}/\Lambda^{4}$  & [-6.27; 5.84] & [-4.27; 3.84]  & [-3.62; 3.20] & [-2.81; 2.38]   \\
\hline
$f_{M2}/\Lambda^{4}$  & [-5.30; 5.52] $\times 10^{-1}$ & [-3.51; 3.73]$ \times 10^{-1}$  & [-2.94; 3.15] $ \times 10^{-1}$ & [-2.21; 2.42]$\times 10^{-1}$   \\
\hline
$f_{M3}/\Lambda^{4}$  & [-9.57; 9.08] $\times 10^{-1}$ & [-6.48; 6.00]$ \times 10^{-1}$  & [-5.49; 5.00] $ \times 10^{-1}$ & [-4.23; 3.75]$\times 10^{-1}$   \\
\hline
$f_{M4}/\Lambda^{4}$  & [-1.96; 1.95] & [-1.31; 1.30]  & [-1.11; 1.10] & [-8.41; 8.31]$\times 10^{-1}$   \\
\hline
$f_{M5}/\Lambda^{4}$  & [-3.13; 3.59] & [-2.02; 2.49]  & [-1.67; 2.13] & [-1.22; 1.68]   \\
\hline
$f_{M7}/\Lambda^{4}$  & [-1.21; 1.24]$\times 10^{1}$    & [-8.02; 8.32]   & [-6.72; 7.02]   & [-5.07; 5.37]    \\
\hline\hline
$f_{T0}/\Lambda^{4}$  &  [-2.33; 2.62] $\times 10^{-3}$ & [-1.52; 1.80] $\times 10^{-3}$ & [-1.26; 1.54] $\times 10^{-3}$ & [-0.92; 1.21] $\times 10^{-3}$ \\
\hline
$f_{T1}/\Lambda^{4}$  & [-3.61;3.45] $\times 10^{-3}$  & [-2.44; 2.28] $\times 10^{-3}$ & [-2.07; 1.91] $\times 10^{-3}$ & [-1.59; 1.43] $\times 10^{-3}$  \\
\hline
$f_{T2}/\Lambda^{4}$  &  [-7.30; 7.11] $\times 10^{-3}$ & [-4.91; 4.73] $\times 10^{-3}$ & [-4.14; 3.96] $\times 10^{-3}$ & [-3.17; 2.99] $\times 10^{-3}$ \\
\hline
$f_{T5}/\Lambda^{4}$  & [-7.40; 7.65] $\times 10^{-4}$  & [-4.91; 5.16] $\times 10^{-4}$ & [-4.11; 4.36] $\times 10^{-4}$ & [-3.09; 3.34] $\times 10^{-4}$  \\
\hline
$f_{T6}/\Lambda^{4}$  & [-1.10; 1.06] $\times 10^{-3}$  & [-7.41; 7.04] $\times 10^{-4}$ & [-6.26; 5.89] $\times 10^{-4}$ & [-4.80; 4.44] $\times 10^{-4}$ \\
\hline
$f_{T7}/\Lambda^{4}$  & [-2.10; 2.30] $\times 10^{-3}$  & [-1.37; 1.58] $\times 10^{-3}$ & [-1.14; 1.34] $\times 10^{-3}$ & [-0.85; 1.05] $\times 10^{-3}$ \\
\hline\hline
\multicolumn{5}{|c|}{ $\sqrt{s}= 100$ TeV, \hspace{5mm} $\delta_{sys}=3\%$,  \hspace{5mm} $95\%$ C.L. } \\
\hline
\cline{1-5}
$f_{M0}/\Lambda^{4}$  & [-3.96;3.50] & [-3.10;2.65]  & [-2.92;2.46]  & [-2.77;2.32] \\
\hline
$f_{M1}/\Lambda^{4}$  & [-6.59;6.17] & [-5.12;4.70]  & [-4.81;4.38] & [-4.56;4.13]  \\
\hline
$f_{M2}/\Lambda^{4}$  & [-5.59;5.81] $\times 10^{-1}$ & [-4.28;4.50]$ \times 10^{-1}$  & [-4.00;4.22] $ \times 10^{-1}$ & [-3.77;3.99] $\times 10^{-1}$  \\
\hline
$f_{M3}/\Lambda^{4}$  & [-10.0;9.58] $\times 10^{-1}$ & [-7.80;7.32]$ \times 10^{-1}$  & [-7.32;6.83] $ \times 10^{-1}$ & [-6.93;6.45] $\times 10^{-1}$  \\
\hline
$f_{M4}/\Lambda^{4}$  & [-2.07;2.06] & [-1.59;1.58]  & [-1.49;1.48] & [-1.41;1.40]  \\
\hline
$f_{M5}/\Lambda^{4}$  & [-3.31;3.77] & [-2.50;2.96]  & [-2.32;2.79] & [-2.19;2.65]  \\
\hline
$f_{M7}/\Lambda^{4}$  & [-1.27;1.30] $\times 10^{1}$ & [-9.75;10.05]  & [-9.12;9.42] & [-8.61;8.91] \\
\hline\hline
$f_{T0}/\Lambda^{4}$  & [-2.46;2.75] $\times 10^{-3}$ & [-1.86;1.15] $\times 10^{-3}$  & [-1.74;2.02] $ \times 10^{-3}$ & [-1.63;1.92] $\times 10^{-3}$ \\
\hline
$f_{T1}/\Lambda^{4}$  & [-3.80;3.64] $\times 10^{-3}$ & [-2.94;2.78] $\times 10^{-3}$  & [-2.76;2.60] $ \times 10^{-3}$ & [-2.61;2.45] $\times 10^{-3}$  \\
\hline
$f_{T2}/\Lambda^{4}$  & [-7.68;7.50] $\times 10^{-3}$ & [-5.93;5.74] $\times 10^{-3}$  & [-5.55;5.37] $ \times 10^{-3}$ & [-5.25;5.07] $\times 10^{-3}$ \\
\hline
$f_{T5}/\Lambda^{4}$  & [-7.81;8.06] $\times 10^{-4}$ & [-5.98;6.23] $\times 10^{-4}$  & [-5.60;5.85] $\times 10^{-4}$    & [-5.29;5.54] $\times 10^{-4}$  \\
\hline
$f_{T6}/\Lambda^{4}$  & [-1.16;1.12]    $\times 10^{-3}$             & [-8.95;8.58] $\times 10^{-4}$     & [-8.39;8.03] $\times 10^{-4}$   & [-7.95;7.58] $\times 10^{-4}$  \\
\hline
$f_{T7}/\Lambda^{4}$  & [-2.22;2.42] $\times 10^{-3}$ & [-1.69;1.89] $\times 10^{-3}$  & [-1.57;1.78] $ \times 10^{-3}$ & [-1.48;1.69] $\times 10^{-3}$ \\
\hline
\end{tabular}
\end{table}

%Table 2
\begin{table}[h]
\caption{Sensitivity at $95\%$ C.L. on the aQGC $W^+W^-\gamma\gamma$ of the $pp \to W\gamma\gamma$ signal for $\sqrt{s}=100$ TeV,
${\cal L}=1, 5, 10,30$ ${\rm ab^{-1}}$ and $\delta_{sys}= 5\%, 10\%$ at the FCC-hh.}
\begin{tabular}{|c|c|c|c|c|c|}
\hline
\multicolumn{5}{|c|}{Leptonic channel} \\
\hline
\multicolumn{5}{|c|}{$\sqrt{s}= 100$ TeV, \hspace{5mm} $\delta_{sys}=5\%$,  \hspace{5mm} $95\%$ C.L.}\\
\hline
Couplings (TeV$^{-4}$) & 1 ab$^{-1}$ & 5 ab$^{-1}$ & 10 ab$^{-1}$ & 30 ab$^{-1}$ \\
\hline
$f_{M0}/\Lambda^{4}$  & [-4.24; 3.78] & [-3.62; 3.17] & [-3.52; 3.06] & [-3.44; 2.99] \\
\hline
$f_{M1}/\Lambda^{4}$  & [-7.07; 6.64] & [-6.02; 5.59]  & [-5.84; 5.41] & [-5.71; 5.28]   \\
\hline
$f_{M2}/\Lambda^{4}$  & [-6.02; 6.24] $\times 10^{-1}$ & [-5.08; 5.29]$ \times 10^{-1}$  & [-4.91; 5.13] $ \times 10^{-1}$ & [-4.80; 5.02]$\times 10^{-1}$   \\
\hline
$f_{M3}/\Lambda^{4}$  & [-10.8; 10.3] $\times 10^{-1}$ & [-9.17; 8.69]$ \times 10^{-1}$  & [-8.90; 8.41] $ \times 10^{-1}$ & [-8.70; 8.21]$\times 10^{-1}$   \\
\hline
$f_{M4}/\Lambda^{4}$  & [-2.22; 2.21] & [-1.88; 1.87]  & [-1.82; 1.81] & [-1.78; 1.77]   \\
\hline
$f_{M5}/\Lambda^{4}$  & [-3.57; 4.04] & [-2.99; 3.45]  & [-2.89; 3.35] & [-2.82; 3.28]   \\
\hline
$f_{M7}/\Lambda^{4}$  & [-1.37; 1.40]$\times 10^{1}$    & [-1.16; 1.19] $\times 10^{1}$  & [-1.12; 1.15] $\times 10^{1}$  & [-1.09; 1.12] $\times 10^{1}$   \\
\hline\hline
$f_{T0}/\Lambda^{4}$  &  [-2.66; 2.94] $\times 10^{-3}$ & [-2.23; 2.51] $\times 10^{-3}$ & [-2.15; 2.44] $\times 10^{-3}$ & [-2.10; 2.38] $\times 10^{-3}$ \\
\hline
$f_{T1}/\Lambda^{4}$  & [-4.08;3.92] $\times 10^{-3}$  & [-3.46; 3.30] $\times 10^{-3}$ & [-3.36; 3.20] $\times 10^{-3}$ & [-3.28; 3.12] $\times 10^{-3}$  \\
\hline
$f_{T2}/\Lambda^{4}$  &  [-8.24; 8.06] $\times 10^{-3}$ & [-6.99; 6.80] $\times 10^{-3}$ & [-6.77; 6.59] $\times 10^{-3}$ & [-6.62; 6.44] $\times 10^{-3}$ \\
\hline
$f_{T5}/\Lambda^{4}$  & [-8.41; 8.66] $\times 10^{-4}$  & [-7.10; 7.35] $\times 10^{-4}$ & [-6.88; 7.13] $\times 10^{-4}$ & [-6.72; 6.97] $\times 10^{-4}$  \\
\hline
$f_{T6}/\Lambda^{4}$  & [-1.24; 1.21] $\times 10^{-3}$  & [-1.06; 1.02] $\times 10^{-3}$ & [-1.02; 0.99] $\times 10^{-3}$ & [-1.00; 0.96] $\times 10^{-3}$ \\
\hline
$f_{T7}/\Lambda^{4}$  & [-2.40; 2.60] $\times 10^{-3}$  & [-2.01; 2.22] $\times 10^{-3}$ & [-1.95; 2.15] $\times 10^{-3}$ & [-1.90; 2.10] $\times 10^{-3}$ \\
\hline\hline
\multicolumn{5}{|c|}{ $\sqrt{s}= 100$ TeV, \hspace{5mm} $\delta_{sys}=10\%$,  \hspace{5mm} $95\%$ C.L. } \\
\hline
\cline{1-5}
$f_{M0}/\Lambda^{4}$  & [-5.09;4.64] & [-4.79;4.34]  & [-4.75;4.30] & [-4.72;4.27] \\
\hline
$f_{M1}/\Lambda^{4}$  & [-8.54;8.11] & [-8.03;7.60]  & [-7.95;7.53] & [-7.91;7.48]  \\
\hline
$f_{M2}/\Lambda^{4}$  & [-7.33;7.55] $\times 10^{-1}$ & [-6.87;7.09]$ \times 10^{-1}$  & [-6.81;7.03] $ \times 10^{-1}$ & [-6.76;6.98] $\times 10^{-1}$  \\
\hline
$f_{M3}/\Lambda^{4}$  & [-1.31;1.26] & [-1.23;1.18]  & [-1.22;1.17] & [-1.21;1.16]  \\
\hline
$f_{M4}/\Lambda^{4}$  & [-2.69;2.68] & [-2.53;2.52]  & [-2.51;2.50] & [-2.49;2.48]  \\
\hline
$f_{M5}/\Lambda^{4}$  & [-4.39;4.85] & [-4.10;4.57]  & [-4.06;4.53] & [-4.04;4.50]  \\
\hline
$f_{M7}/\Lambda^{4}$  & [-1.66;1.69] $\times 10^{1}$ & [-1.56;1.59]$ \times 10^{1}$  & [-1.55;1.58] $ \times 10^{1}$ & [-1.54;1.57] $\times 10^{1}$  \\
\hline\hline
$f_{T0}/\Lambda^{4}$  & [-3.26;3.54] $\times 10^{-3}$ & [-3.05;3.33] $\times 10^{-3}$  & [-3.02;3.30] $ \times 10^{-3}$ & [-3.00;3.28] $\times 10^{-3}$ \\
\hline
$f_{T1}/\Lambda^{4}$  & [-4.93;4.77] $\times 10^{-3}$ & [-4.63;4.47] $\times 10^{-3}$  & [-4.59;4.43] $ \times 10^{-3}$ & [-4.56;4.40] $\times 10^{-3}$  \\
\hline
$f_{T2}/\Lambda^{4}$  & [-9.99;9.81] $\times 10^{-3}$ & [-9.38;9.20] $\times 10^{-3}$  & [-9.29;9.11] $ \times 10^{-3}$ & [-9.23;9.05] $\times 10^{-3}$ \\
\hline
$f_{T5}/\Lambda^{4}$  & [-1.02;1.05] $\times 10^{-3}$ & [-9.61;9.86] $\times 10^{-4}$ & [-9.52;9.77] $\times 10^{-4}$          & [-9.46;9.71] $\times 10^{-4}$  \\
\hline
$f_{T6}/\Lambda^{4}$  & [-1.51;1.47]    $\times 10^{-3}$             & [-1.41;1.38] $\times 10^{-3}$                 & [-1.40;1.37] $\times 10^{-3}$                 & [-1.39;1.36] $\times 10^{-3}$  \\
\hline
$f_{T7}/\Lambda^{4}$  & [-2.93;3.13] $\times 10^{-3}$ & [-2.75;2.95] $\times 10^{-3}$  & [-2.72;2.92] $ \times 10^{-3}$ & [-2.70;2.90] $\times 10^{-3}$ \\
\hline
\end{tabular}
\end{table}

%Table 3
\begin{table}[h]
\caption{Sensitivity at $95\%$ C.L. on the aQGC $W^+W^-\gamma\gamma$ of the $pp \to W\gamma\gamma$ signal for $\sqrt{s}=100$ TeV,
${\cal L}=1, 5, 10,30$ ${\rm ab^{-1}}$ and $\delta_{sys}=0\%, 3\%$ at the FCC-hh.}
\begin{tabular}{|c|c|c|c|c|c|}
\hline
\multicolumn{5}{|c|}{Hadronic channel} \\
\hline
\multicolumn{5}{|c|}{ $\sqrt{s}= 100$ TeV, \hspace{5mm} $\delta_{sys}=0\%$,  \hspace{5mm} $95\%$ C.L. }\\
\hline
Couplings (TeV$^{-4}$) & 1 ab$^{-1}$ & 5 ab$^{-1}$ & 10 ab$^{-1}$ & 30 ab$^{-1}$ \\
\hline
$f_{M0}/\Lambda^{4}$  & [-4.03; 3.97] $\times 10^{1}$ & [-2.71; 2.65] $\times 10^{1}$ & [-2.28; 2.22] $\times 10^{1}$ & [-1.74; 1.68] $\times 10^{1}$ \\
\hline
$f_{M1}/\Lambda^{4}$  & [-0.69; 0.59] $\times 10^{2}$ & [-0.48; 0.38]$ \times 10^{2}$  & [-0.41; 0.31] $ \times 10^{2}$ & [-0.33; 0.23] $ \times 10^{2}$   \\
\hline
$f_{M2}/\Lambda^{4}$  & [-5.91; 6.27] & [-3.89; 4.26]  & [-3.25; 3.61] & [-2.43; 2.79]   \\
\hline
$f_{M3}/\Lambda^{4}$  & [-9.70; 9.13] & [-6.59; 6.01]  & [-5.59; 5.01] & [-4.32; 3.74]   \\
\hline
$f_{M4}/\Lambda^{4}$  & [-2.25; 2.17] $\times 10^{1}$ & [-1.52; 1.44]$ \times 10^{1}$  & [-1.28; 1.20] $ \times 10^{1}$ & [-9.85; 9.05]   \\
\hline
$f_{M5}/\Lambda^{4}$  & [-3.43; 3.50] $\times 10^{1}$ & [-2.28; 2.35]$ \times 10^{1}$  & [-1.91; 1.99] $ \times 10^{1}$ & [-1.44; 1.52]$\times 10^{1}$   \\
\hline
$f_{M7}/\Lambda^{4}$  & [-1.17; 1.24]$\times 10^{2}$    & [-0.77; 0.84]$\times 10^{2}$   & [-0.65; 0.71]$\times 10^{2}$   & [-0.48; 0.55]$\times 10^{2}$    \\
\hline
$f_{T0}/\Lambda^{4}$  &  [-5.14; 5.25] $\times 10^{-2}$ & [-3.42; 3.53] $\times 10^{-2}$ & [-2.87; 2.98] $\times 10^{-2}$ & [-2.16; 2.28] $\times 10^{-2}$ \\
\hline\hline
$f_{T1}/\Lambda^{4}$  & [-1.29;0.69] $\times 10^{-1}$  & [-1.00; 0.40] $\times 10^{-1}$ & [-0.91; 0.31] $\times 10^{-1}$ & [-0.80; 0.20] $\times 10^{-1}$  \\
\hline
$f_{T2}/\Lambda^{4}$  &  [-1.44; 0.90] $\times 10^{-1}$ & [-1.08; 0.54] $\times 10^{-1}$ & [-0.97; 0.42] $\times 10^{-1}$ & [-0.83; 0.28] $\times 10^{-1}$ \\
\hline
$f_{T5}/\Lambda^{4}$  & [-1.44; 1.51] $\times 10^{-2}$  & [-0.95; 1.02] $\times 10^{-2}$ & [-0.79; 0.86] $\times 10^{-2}$ & [-0.60; 0.66] $\times 10^{-2}$  \\
\hline
$f_{T6}/\Lambda^{4}$  & [-2.48; 2.28] $\times 10^{-2}$  & [-1.69; 1.49] $\times 10^{-2}$ & [-1.44; 1.24] $\times 10^{-2}$ & [-1.12; 0.92] $\times 10^{-2}$ \\
\hline
$f_{T7}/\Lambda^{4}$  & [-3.41; 3.94] $\times 10^{-2}$  & [-2.20; 2.73] $\times 10^{-2}$ & [-1.82; 2.34] $\times 10^{-2}$ & [-1.33; 1.85] $\times 10^{-2}$ \\
\hline\hline
\multicolumn{5}{|c|}{ $\sqrt{s}= 100$ TeV, \hspace{5mm} $\delta_{sys}=3\%$,  \hspace{5mm} $95\%$ C.L. } \\
\hline
\cline{1-5}
$f_{M0}/\Lambda^{4}$  & [-4.31;4.26] $\times 10^{1}$ & [-3.42;3.36] $\times 10^{1}$  & [-3.24;3.18] $ \times 10^{1}$ & [-3.10;3.04] $\times 10^{1}$ \\
\hline
$f_{M1}/\Lambda^{4}$  & [-7.38;6.36] $\times 10^{1}$ & [-5.95;4.93]$ \times 10^{1}$  & [-5.66;4.65] $ \times 10^{1}$ & [-5.44;4.43] $\times 10^{1}$  \\
\hline
$f_{M2}/\Lambda^{4}$  & [-6.34;6.70] & [-4.98;5.34]  & [-4.71;5.07] & [-4.50;4.86]  \\
\hline
$f_{M3}/\Lambda^{4}$  & [-10.3;9.79] & [-8.26;7.69]  & [-7.84;7.26] & [-7.52;6.94]  \\
\hline
$f_{M4}/\Lambda^{4}$  & [-2.41;2.33] $\times 10^{1}$ & [-1.91;1.83]$ \times 10^{1}$  & [-1.81;1.73] $ \times 10^{1}$ & [-1.74;1.66] $\times 10^{1}$  \\
\hline
$f_{M5}/\Lambda^{4}$  & [-3.67;3.75] $\times 10^{1}$ & [-2.90;2.97]$ \times 10^{1}$  & [-2.74;2.82] $ \times 10^{1}$ & [-2.62;2.70] $\times 10^{1}$  \\
\hline
$f_{M7}/\Lambda^{4}$  & [-1.26;1.32] $\times 10^{2}$ & [-0.99;1.05]$ \times 10^{2}$  & [-9.35;9.99] $ \times 10^{1}$ & [-8.93;9.57] $\times 10^{1}$  \\
\hline\hline
$f_{T0}/\Lambda^{4}$  & [-5.51;5.62] $\times 10^{-2}$ & [-4.35;4.46] $\times 10^{-2}$  & [-4.12;4.23] $ \times 10^{-2}$ & [-3.94;4.06] $\times 10^{-2}$ \\
\hline
$f_{T1}/\Lambda^{4}$  & [-1.35;0.75] $\times 10^{-1}$ & [-1.15;0.55] $\times 10^{-1}$  & [-1.11;0.51] $ \times 10^{-1}$ & [-1.08;0.48] $\times 10^{-1}$  \\
\hline
$f_{T2}/\Lambda^{4}$  & [-1.52;0.98] $\times 10^{-1}$ & [-1.28;0.73] $\times 10^{-1}$  & [-1.23;0.68] $ \times 10^{-1}$ & [-1.19;0.64] $\times 10^{-1}$ \\
\hline
$f_{T5}/\Lambda^{4}$  & [-1.54;1.61] $\times 10^{-2}$ & [-1.21;1.28] $\times 10^{-2}$ & [-1.14;1.21] $\times 10^{-2}$      & [-1.09;1.16] $\times 10^{-2}$  \\
\hline
$f_{T6}/\Lambda^{4}$  & [-2.64;2.44]    $\times 10^{-2}$             & [-2.11;1.91] $\times 10^{-2}$     & [-2.00;1.80] $\times 10^{-2}$     & [-1.92;1.72] $\times 10^{-2}$  \\
\hline
$f_{T7}/\Lambda^{4}$  & [-3.67;4.19] $\times 10^{-2}$ & [-2.85;3.37] $\times 10^{-2}$  & [-2.68;3.21] $ \times 10^{-2}$ & [-2.56;3.08] $\times 10^{-2}$ \\
\hline
\end{tabular}
\end{table}

%Table 4
\begin{table}[h]
\caption{Sensitivity at $95\%$ C.L. on the aQGC $W^+W^-\gamma\gamma$ of the $pp \to W\gamma\gamma$ signal for $\sqrt{s}=100$ TeV,
${\cal L}=1, 5, 10,30$ ${\rm ab^{-1}}$ and $\delta_{sys}=5\%, 10\%$ at the FCC-hh.}
\begin{tabular}{|c|c|c|c|c|c|}
\hline
\multicolumn{5}{|c|}{Hadronic channel} \\
\hline
\multicolumn{5}{|c|}{ $\sqrt{s}= 100$ TeV, \hspace{5mm} $\delta_{sys}=5\%$,  \hspace{5mm} $95\%$ C.L. }\\
\hline
Couplings (TeV$^{-4}$) & 1 ab$^{-1}$ & 5 ab$^{-1}$ & 10 ab$^{-1}$ & 30 ab$^{-1}$ \\
\hline
$f_{M0}/\Lambda^{4}$  & [-4.71; 4.65] $\times 10^{1}$ & [-4.10; 4.04] $\times 10^{1}$ & [-4.00; 3.95] $\times 10^{1}$ & [-3.93; 3.88] $\times 10^{1}$ \\
\hline
$f_{M1}/\Lambda^{4}$  & [-8.01; 7.00] $\times 10^{1}$ & [-7.04; 6.02] $\times 10^{1}$ & [-6.88; 5.87] $\times 10^{1}$ & [-6.77; 5.76] $\times 10^{1}$  \\
\hline
$f_{M2}/\Lambda^{4}$  & [-6.94; 7.31] & [-6.02; 6.38]  & [-5.87; 6.23] & [-5.76; 6.12]   \\
\hline
$f_{M3}/\Lambda^{4}$  & [-1.13; 1.07] $\times 10^{1}$ & [-9.87; 9.29]  & [-9.64; 9.06] & [-9.47; 8.90]   \\
\hline
$f_{M4}/\Lambda^{4}$  & [-2.62; 2.54] $\times 10^{1}$ & [-2.29; 2.21] $\times 10^{1}$ & [-2.23; 2.15] $\times 10^{1}$ & [-2.20; 2.12]$\times 10^{1}$   \\
\hline
$f_{M5}/\Lambda^{4}$  & [-4.02; 4.09] $\times 10^{1}$ & [-3.49; 3.56] $\times 10^{1}$ & [-3.40; 3.48] $\times 10^{1}$ & [-3.34; 3.42] $\times 10^{1}$   \\
\hline
$f_{M7}/\Lambda^{4}$  & [-1.38; 1.44]$\times 10^{2}$    & [-1.19; 1.26] $\times 10^{2}$  & [-1.16; 1.23] $\times 10^{2}$  & [-1.14; 1.21] $\times 10^{2}$   \\
\hline\hline
$f_{T0}/\Lambda^{4}$  &  [-6.03; 6.14] $\times 10^{-2}$ & [-5.24; 5.36] $\times 10^{-2}$ & [-5.11; 5.23] $\times 10^{-2}$ & [-5.02; 5.14] $\times 10^{-2}$ \\
\hline
$f_{T1}/\Lambda^{4}$  & [-1.44;0.84] $\times 10^{-1}$  & [-1.30; 0.71] $\times 10^{-1}$ & [-1.28; 0.68] $\times 10^{-1}$ & [-1.27; 0.67] $\times 10^{-1}$  \\
\hline
$f_{T2}/\Lambda^{4}$  &  [-1.63; 1.09] $\times 10^{-1}$ & [-1.46; 0.92] $\times 10^{-1}$ & [-1.44; 0.89] $\times 10^{-1}$ & [-1.42; 0.87] $\times 10^{-1}$ \\
\hline
$f_{T5}/\Lambda^{4}$  & [-1.68; 1.75] $\times 10^{-2}$  & [-1.46; 1.53] $\times 10^{-2}$ & [-1.42; 1.49] $\times 10^{-2}$ & [-1.40; 1.47] $\times 10^{-2}$  \\
\hline
$f_{T6}/\Lambda^{4}$  & [-2.88; 2.68] $\times 10^{-2}$  & [-2.51; 2.31] $\times 10^{-2}$ & [-2.46; 2.25] $\times 10^{-2}$ & [-2.41; 2.21] $\times 10^{-2}$ \\
\hline
$f_{T7}/\Lambda^{4}$  & [-4.03; 4.55] $\times 10^{-2}$  & [-3.47; 3.99] $\times 10^{-2}$ & [-3.38; 3.90] $\times 10^{-2}$ & [-3.31; 3.84] $\times 10^{-2}$ \\
\hline\hline
\multicolumn{5}{|c|}{ $\sqrt{s}= 100$ TeV, \hspace{5mm} $\delta_{sys}=10\%$,  \hspace{5mm} $95\%$ C.L. } \\
\hline
\cline{1-5}
$f_{M0}/\Lambda^{4}$  & [-5.86;5.80] $\times 10^{1}$ & [-5.58;5.52] $\times 10^{1}$  & [-5.54;5.48] $ \times 10^{1}$ & [-5.51;5.45] $\times 10^{1}$ \\
\hline
$f_{M1}/\Lambda^{4}$  & [-9.83;8.82] $\times 10^{1}$ & [-9.39;8.38]$ \times 10^{1}$  & [-9.33;8.32] $ \times 10^{1}$ & [-9.29;8.27] $\times 10^{1}$  \\
\hline
$f_{M2}/\Lambda^{4}$  & [-8.68;9.05] & [-8.26;8.62]  & [-8.20;8.56] & [-8.16;8.52]  \\
\hline
$f_{M3}/\Lambda^{4}$  & [-1.40;1.34] $\times 10^{1}$ & [-1.33;1.28]$ \times 10^{1}$  & [-1.32;1.27] $ \times 10^{1}$ & [-1.32;1.26] $\times 10^{1}$  \\
\hline
$f_{M4}/\Lambda^{4}$  & [-3.26;3.18] $\times 10^{1}$ & [-3.10;3.02]$ \times 10^{1}$  & [-3.08;3.00] $ \times 10^{1}$ & [-3.07;2.99] $\times 10^{1}$  \\
\hline
$f_{M5}/\Lambda^{4}$  & [-5.01;5.08] $\times 10^{1}$ & [-4.76;4.84]$ \times 10^{1}$  & [-4.73;4.81] $ \times 10^{1}$ & [-4.71;4.78] $\times 10^{1}$  \\
\hline
$f_{M7}/\Lambda^{4}$  & [-1.72;1.79] $\times 10^{2}$ & [-1.64;1.70]$ \times 10^{2}$  & [-1.63;1.69] $ \times 10^{2}$ & [-1.62;1.68] $\times 10^{2}$  \\
\hline\hline
$f_{T0}/\Lambda^{4}$  & [-7.52;7.64] $\times 10^{-2}$ & [-7.16;7.28] $\times 10^{-2}$  & [-7.11;7.23] $ \times 10^{-2}$ & [-7.08;7.19] $\times 10^{-2}$ \\
\hline
$f_{T1}/\Lambda^{4}$  & [-1.70;1.10] $\times 10^{-1}$ & [-1.64;1.04] $\times 10^{-1}$  & [-1.63;1.03] $ \times 10^{-1}$ & [-1.62;1.02] $\times 10^{-1}$  \\
\hline
$f_{T2}/\Lambda^{4}$  & [-1.95;1.41] $\times 10^{-1}$ & [-1.88;1.33] $\times 10^{-1}$  & [-1.87;1.32] $ \times 10^{-1}$ & [-1.86;1.31] $\times 10^{-1}$ \\
\hline
$f_{T5}/\Lambda^{4}$  & [-2.10;2.17] $\times 10^{-2}$ & [-2.00;2.07] $\times 10^{-2}$ & [-1.98;2.05]    $\times 10^{-2}$  & [-1.97;2.04] $\times 10^{-2}$  \\
\hline
$f_{T6}/\Lambda^{4}$  & [-3.55;3.35] $\times 10^{-2}$                 & [-3.38;3.18] $\times 10^{-2}$         & [-3.36;3.16] $\times 10^{-2}$    & [-3.35;3.14] $\times 10^{-2}$  \\
\hline
$f_{T7}/\Lambda^{4}$  & [-5.07;5.59] $\times 10^{-2}$ & [-4.81;5.33] $\times 10^{-2}$  & [-4.77;5.30] $ \times 10^{-2}$ & [-4.75;5.28] $\times 10^{-2}$ \\
\hline
\end{tabular}
\end{table}

Searches for processes containing aQGC have been performed by previous and present experiments. For instance, in Refs. \cite{OPAL-PRD70,OPAL-PLB580,DELPHI-EPJC31,L3-PLB527} the aQGC were examined using the processes $e^+e^- \rightarrow WW\gamma$,
$e^+e^- \rightarrow \nu\nu\gamma\gamma$ and $e^+e^- \rightarrow  qq\gamma\gamma$ by the OPAL, DELPHI and L3 Collaborations;
$p\bar p \to p W^+W^-\bar p \to p e^+\nu e^- \bar \nu p$ by the D0 Collaboration \cite{D0-PRD88}; $p p \to WV\gamma \to l\nu qq\gamma$
\cite{CMS-EPJC73} and $p p \to pW^+W^-p \to p e^\pm\nu \mu^\mp\nu p$ \cite{CMS-JHEP08,CMS-JHEP10} by the CMS Collaboration; $p p (\gamma\gamma)
\to p W^+ W^- p \to \to p e^\pm\nu \mu^\mp\nu p$ \cite{ATLAS-PRD94} and  $p p \to p W\gamma\gamma p \to p l \nu \gamma\gamma p$
\cite{ATLAS-PRL2015} by the ATLAS Collaboration. These results are used to set limits on corresponding aQGC with at least
one photon involved.

Recently, many phenomenological studies available in the literature have extensively investigated the anomalous $\frac{f_{M, i}}{\Lambda^4}$
and $\frac{f_{T, j}}{\Lambda^4}$ couplings in different contexts and in various future high energy colliders, including hadron-hadron,
lepton-lepton and lepton-hadron colliders, see, e.g. Refs. \cite{ATLAS-PRL2015,ATLAS-PRD2016,ALEPH-Barate,DELPHI-Abreu,L3-Acciarri,OPAL-Abbiendi,CDF-Gounder,D0-Abbott,CMS-Chatrchyan,ATLAS-Aaboud,
LHeC-FCC-he-WWgg-Ari1,LHeC-FCC-he-WWgg-Ari2,Stirling,Leil,Bervan,Chong,Koksal,Stirling1,Atag,Eboli1,Sahin,Koksal1,Chapon,Koksal2,
Senol,Koksal3,Yang,Eboli2,Eboli4,Bell,Ahmadov,Schonherr,Wen,Ye,Perez,Sahin1,Senol1,Baldenegro,Fichet,Pierzchala,Gutierrez,Belanger,Aaboud,
Eboli,Eboli3,Gutierrez-EPJC81-2021} for the most recent reviews. The limit values on the anomalous $\frac{f_{M, i}}{\Lambda^4}$ and
$\frac{f_{T, j}}{\Lambda^4}$ couplings obtained by these studies are of the order of ${\cal O}(10^{-1}-10^1)$. These limit values are weaker
than the bounds obtained in the present study (see Tables II-V).

\section{Conclusions}

In this work, we have calculated the total cross-section of the $pp \to W\gamma\gamma$ process, as well as the prospects
for discovering the anomalous $\frac{f_{M, i}}{\Lambda^4}$ and $\frac{f_{T, j}}{\Lambda^4}$ couplings at the future
FCC-hh with $\sqrt{s}=100$ TeV and ${\cal L}=1, 5, 10, 30$ ${\rm ab^{-1}}$. Furthermore, we considering the systematic
uncertainties of $\delta_{sys}=0\%, 3\%, 5\%, 10\%$. The systematic uncertainties should be taken into account and
discussed in order to quantify the impact on the total cross-section, as well as in the anomalous $WW\gamma\gamma$
couplings in order to make more realistic predictions.

We have performed a simulation for the signals and the relevant SM backgrounds based on two separate cut selections.
The selected cuts for the leptonic and hadronic decay channels of the $W$-boson. We have obtained $95\%$ C.L. sensitivities on the
$\frac{f_{M, i}}{\Lambda^4}$ and $\frac{f_{T, j}}{\Lambda^4}$ parameters. The sensitivity on the anomalous $\frac{f_{M, i}}{\Lambda^4}$
and $\frac{f_{T, j}}{\Lambda^4}$ couplings projected in our study, which are summarised in Tables II-V are expected
to be of ${\cal O}(10^{-4}-10^1)$ for $\delta_{sys}=0\%$. It is interesting to observe that for $\delta_{sys}=3\%, 5\%, 10\%$ the Wilson
coefficients do not degrade substantially and become ${\cal O}(3-4)$ weaker than those obtained with $\delta_{sys}=0\%$. Our results
are compared with those obtained by other experimental and phenomenological groups, which are of ${\cal O}(10^{-1}-10^1)$. Clearly,
our results show a better sensitivity concerning those reported in the literature.

In summary, from our numerical results obtained, we have reached the following conclusions. These new results represent potential
projections for the FCC-hh sensitivity on the aQGC. These limits are nearly three or four orders of magnitude stronger than the
current experimental results obtained by the ATLAS Collaboration from LHC runs at 8 TeV \cite{ATLAS-PRL2015,ATLAS-PRD2016,CMS-JHEP08,CMS-JHEP06-2017,ATLAS-PRD94-2016,CMS-JHEP06-2020}, as well as of
the CMS Collaboration at $\sqrt{s}$=13 TeV \cite{CMS-2105.12780,PLB811-2020} and nearly two or three orders of magnitude
stronger than the current phenomenological results. Therefore, we expect that the signatures studied here to provide
competitive complementary information for detecting the aQGC $WW\gamma\gamma$ at future hadronic colliders such as the FCC-hh
at the CERN.

%\newpage

\vspace{1cm}

\begin{center}
{\bf Acknowledgments}
\end{center}

A. G. R. and M. A. H. R. thank SNI and PROFEXCE (M\'exico). The numerical calculations reported in this paper were fully performed at TUBITAK ULAKBIM, High Performance and Grid Computing Center (TRUBA resources).

\vspace{2cm}

%\newpage

\newpage

\begin{figure}[t]
\centerline{\scalebox{0.5}{\includegraphics{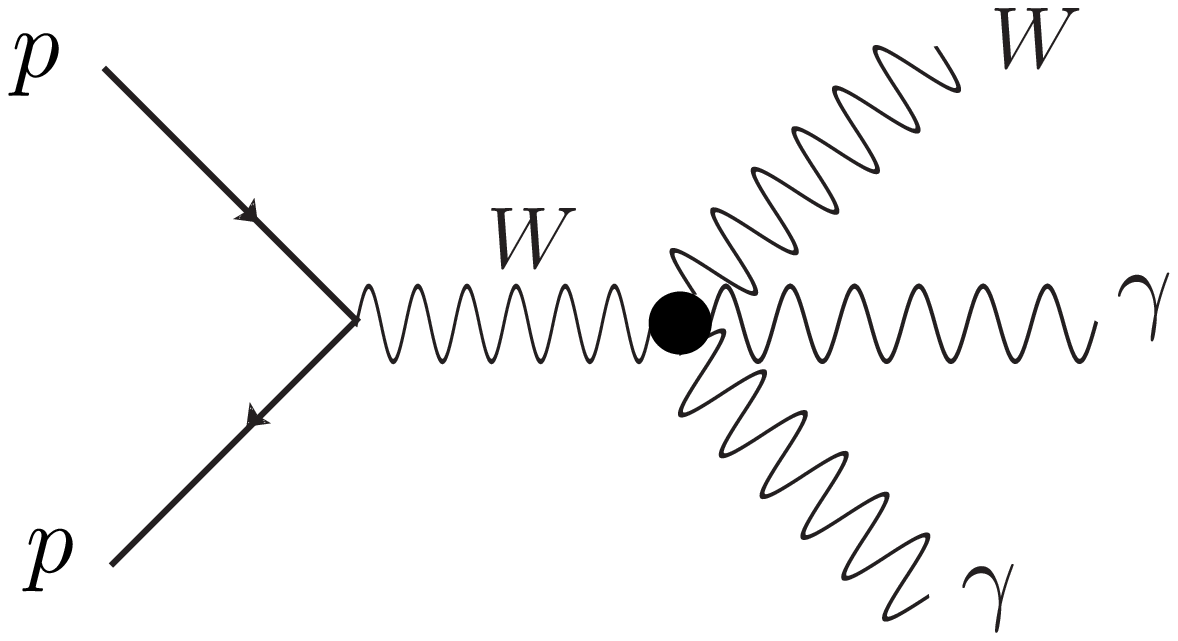}}}
\caption{ \label{fig:gamma1} Feynman diagram for the signal process $pp \to W\gamma\gamma$.
New physics (represented by a black circle) in the electroweak sector can modify the quartic gauge couplings.}
\label{Fig.1}
\end{figure}

\begin{figure}[t]
\centerline{\scalebox{0.5}{\includegraphics{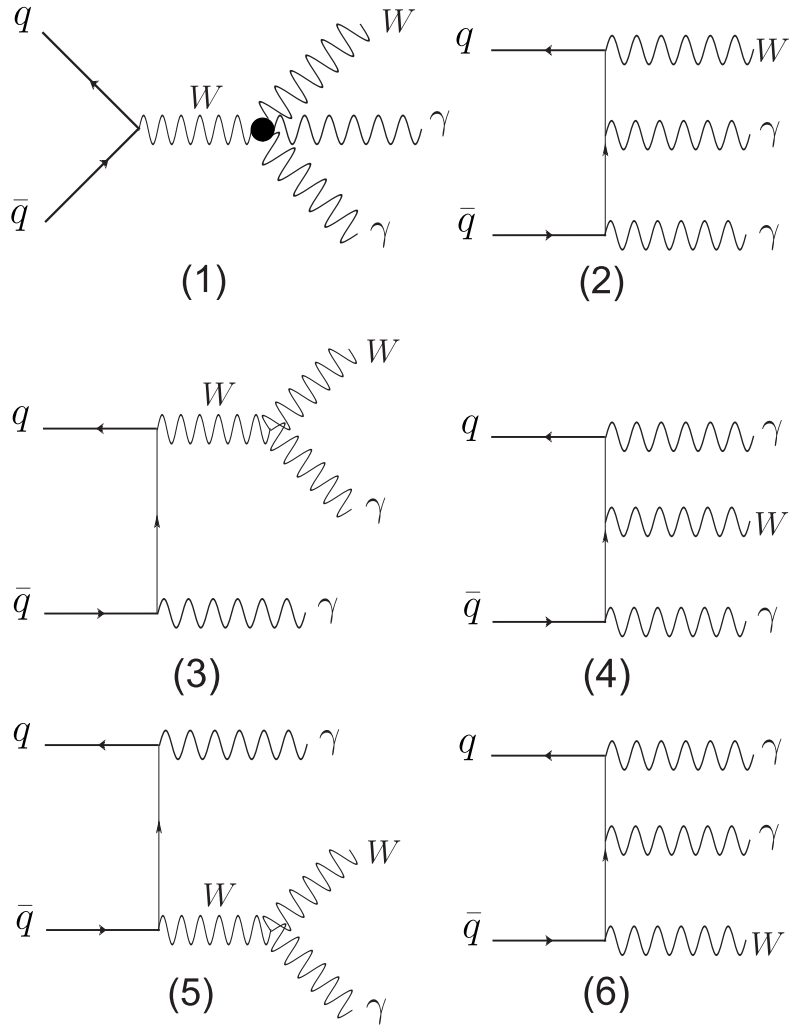}}}
\caption{ \label{fig:gamma2} Representative Feynman diagrams contributing to the subprocess $q\bar{q} \to W\gamma \gamma$.}
\label{Fig.2}
\end{figure}

\begin{figure}[ht]
  \begin{subfigure}[b]{0.5\linewidth}
    \centering
    \includegraphics[width=1.00\linewidth]{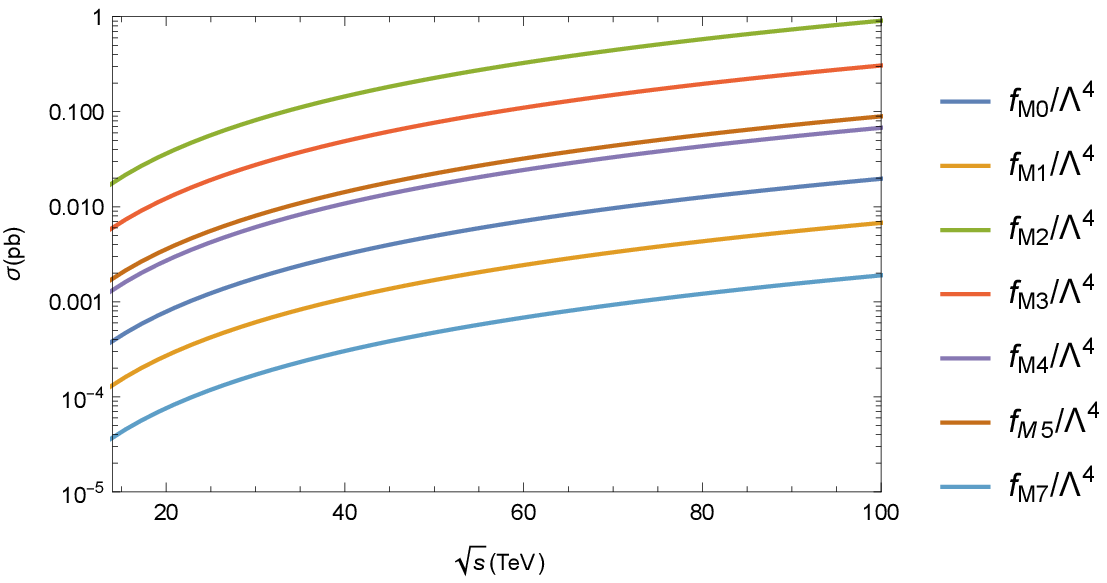}
    \caption{Leptonic decay}
    \label{fig8:a}
    \vspace{4ex}
  \end{subfigure}%%
  \begin{subfigure}[b]{0.5\linewidth}
    \centering
    \includegraphics[width=1.00\linewidth]{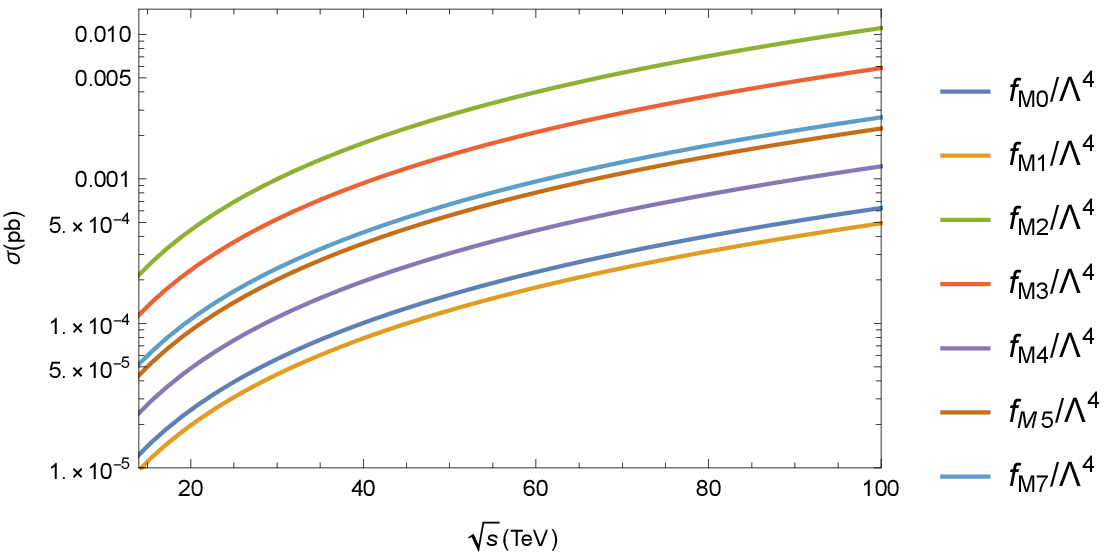}
    \caption{Hadronic decay}
    \label{fig8:b}
    \vspace{4ex}
  \end{subfigure}
  \caption{ a) For leptonic channel, the expected production cross-sections for the $pp \to W\gamma\gamma$
            process as a function of the center-of-mass energy for the anomalous $f_{M,i}/\Lambda^4$ couplings with $j=0,1,2,3,4,5,7$.
            b) Same as in a), but for the hadronic channel.}
  \label{Fig.4}
\end{figure}

\begin{figure}[ht]
  \begin{subfigure}[b]{0.5\linewidth}
    \centering
    \includegraphics[width=1.00\linewidth]{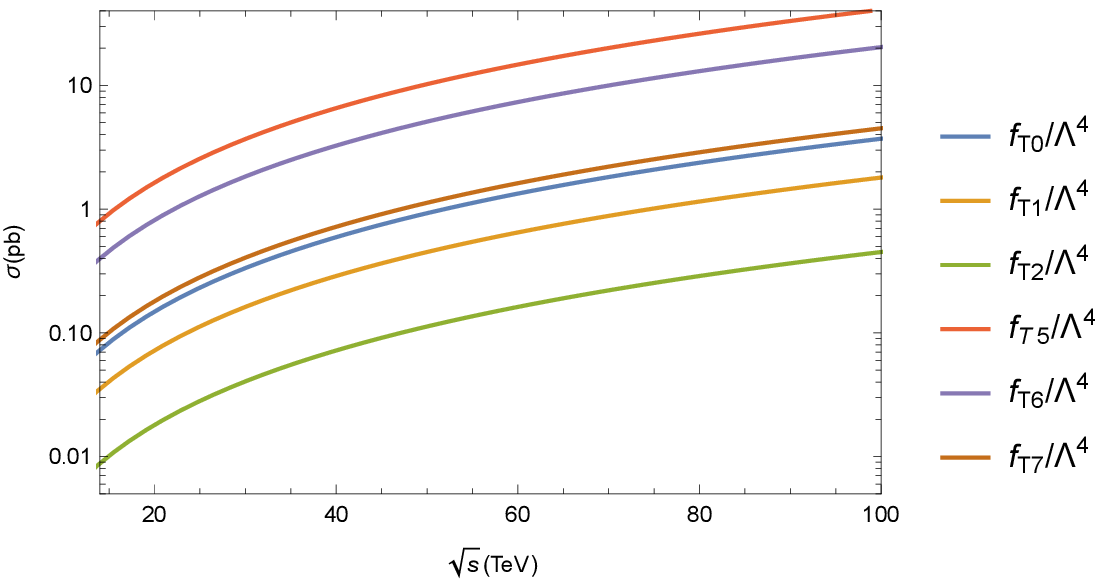}
    \caption{Leptonic decay}
    \label{fig8:a}
    \vspace{4ex}
  \end{subfigure}%%
  \begin{subfigure}[b]{0.5\linewidth}
    \centering
    \includegraphics[width=1.00\linewidth]{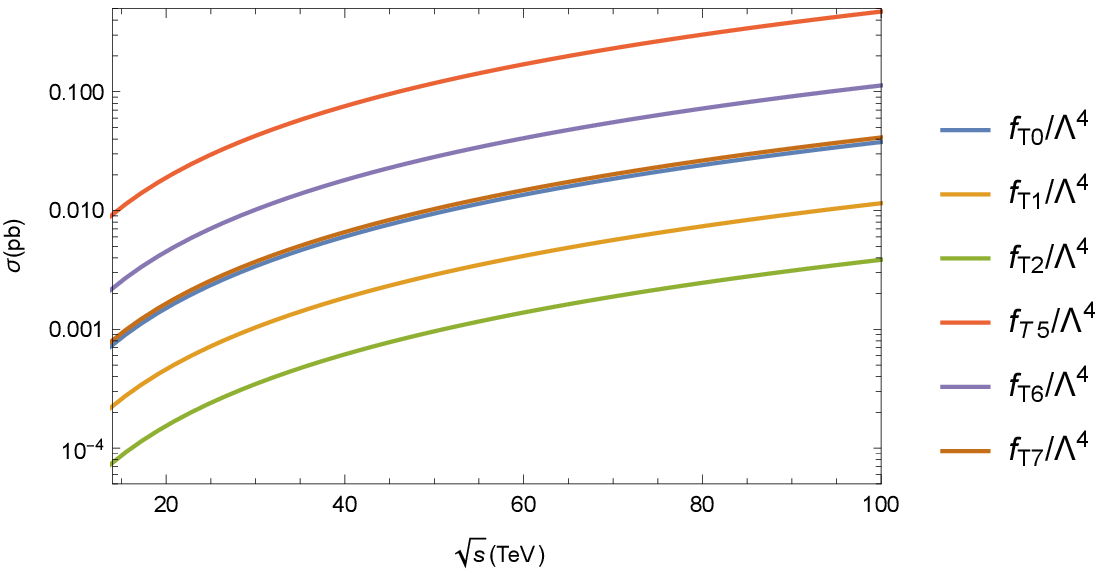}
    \caption{Hadronic decay}
    \label{fig8:b}
    \vspace{4ex}
  \end{subfigure}
  \caption{ a) For leptonic channel, the expected production cross-sections for the $pp \to W\gamma\gamma$
            process as a function of the center-of-mass energy for the anomalous $f_{T,j}/\Lambda^4$ couplings with $j=0,1,2,5,6,7$.
            b) Same as in a), but for the hadronic channel.}
  \label{Fig.4}
\end{figure}

\begin{figure}[t]
\centerline{\scalebox{1.0}{\includegraphics{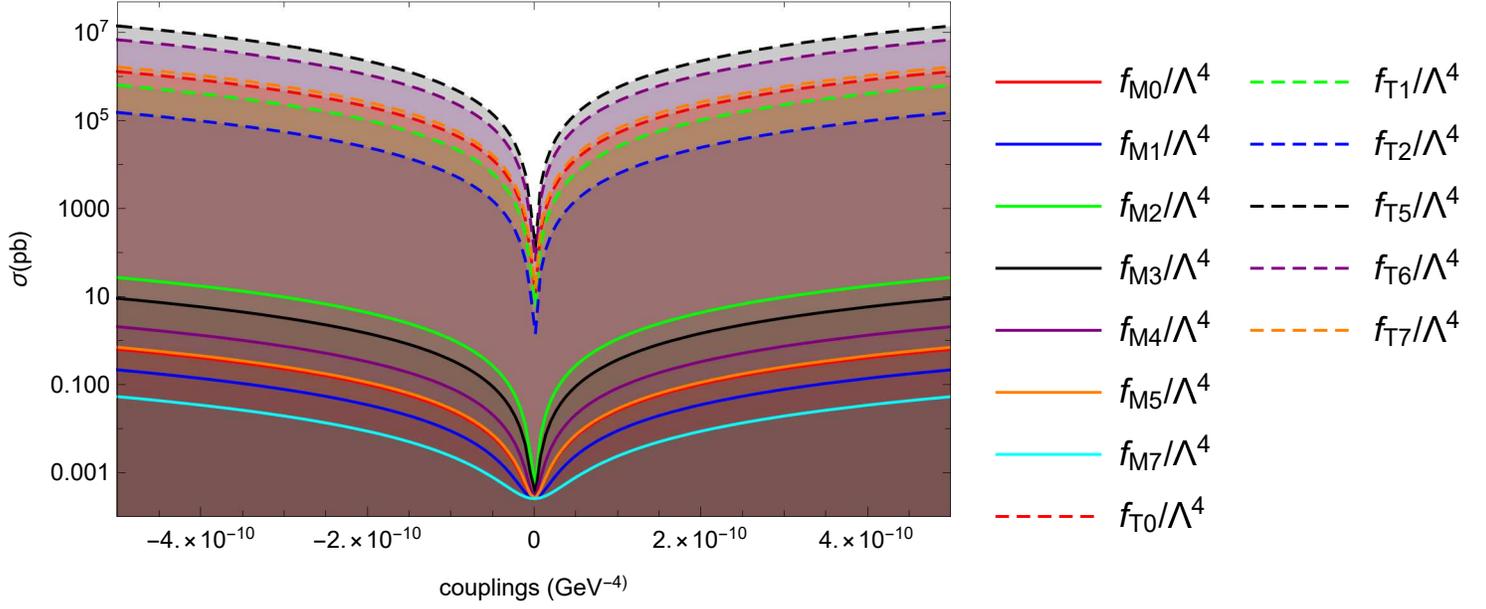}}}
\caption{ \label{fig:gamma1} For leptonic channel, the expected production cross-sections for the $pp \to W\gamma\gamma$
process as a function of the anomalous couplings for the center-of-mass energy of $\sqrt{s}=100$ TeV at the FCC-hh.}
\label{Fig.1}
\end{figure}

\begin{figure}[t]
\centerline{\scalebox{1.0}{\includegraphics{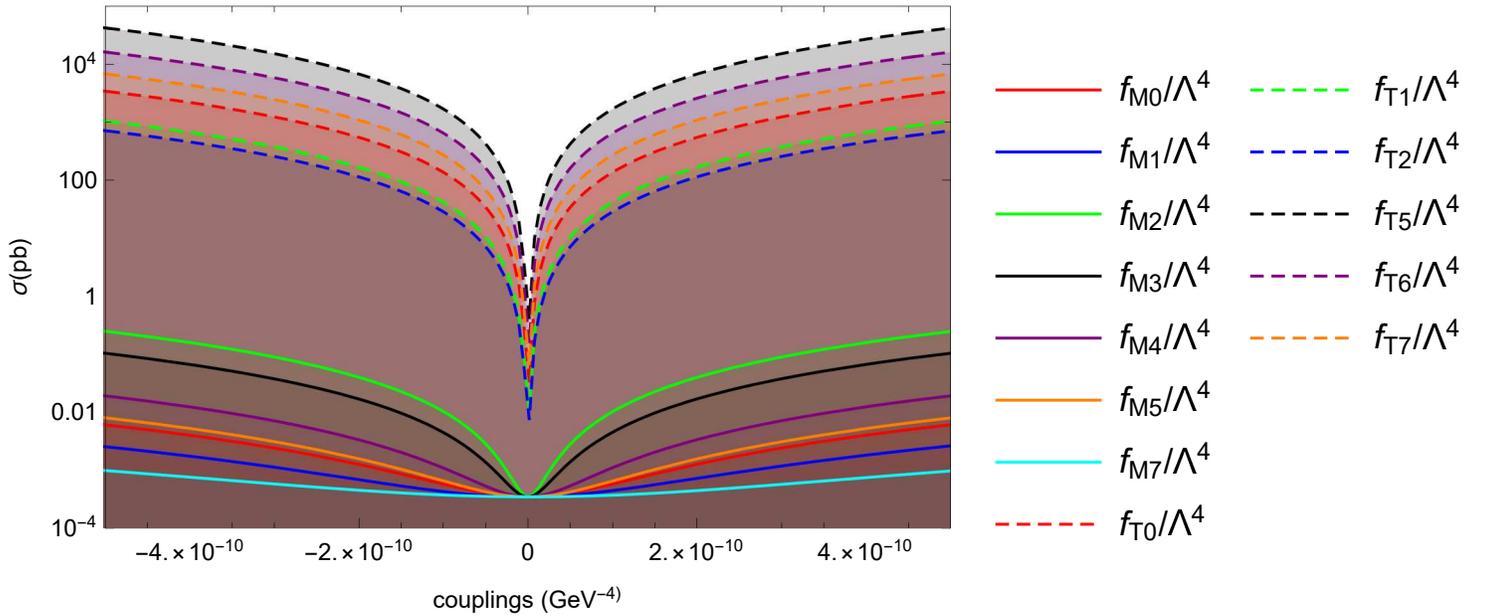}}}
\caption{ \label{fig:gamma2} Same as in Fig. 3, but for the hadronic channel.}
\label{Fig.2}
\end{figure}

\begin{figure}[ht]
  \begin{subfigure}[b]{0.5\linewidth}
    \centering
    \includegraphics[width=1.00\linewidth]{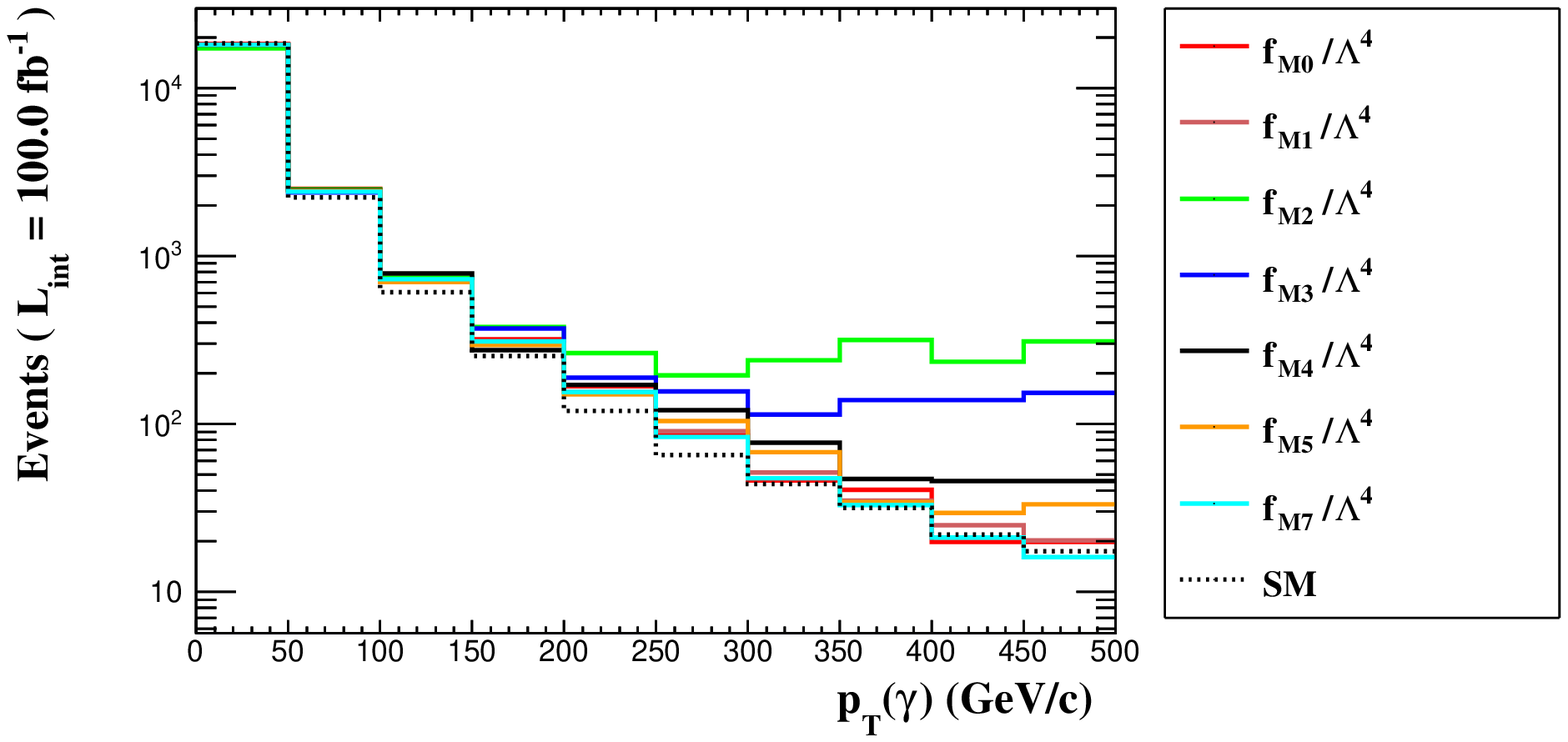}
    \caption{Leptonic decay}
    \label{fig8:a}
    \vspace{4ex}
  \end{subfigure}%%
  \begin{subfigure}[b]{0.5\linewidth}
    \centering
    \includegraphics[width=1.00\linewidth]{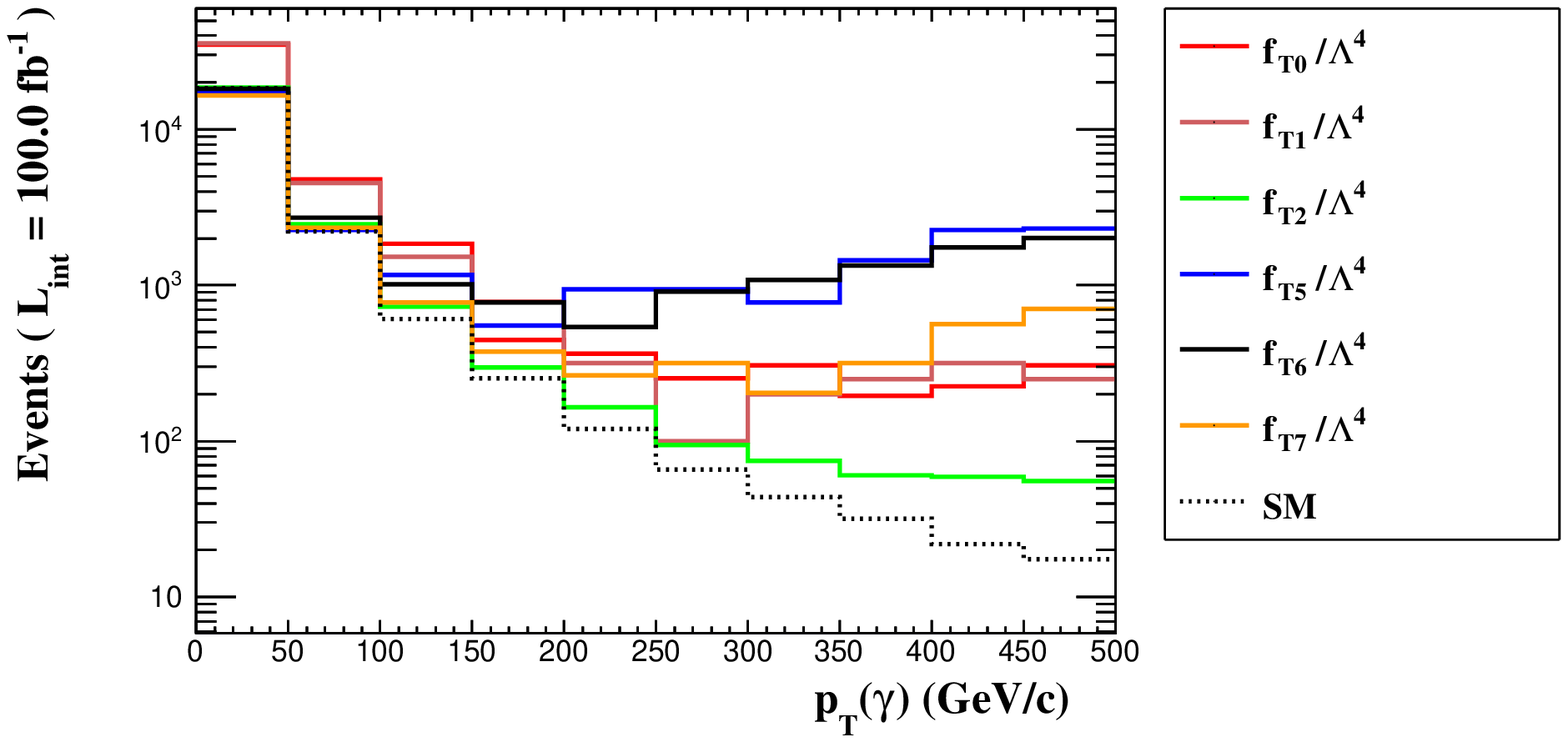}
    \caption{Leptonic decay}
    \label{fig8:b}
    \vspace{4ex}
  \end{subfigure}
  \caption{The number of expected events as a function of the $p^{\gamma}_T$
photon transverse momentum for the $pp \to W\gamma\gamma$ signal and
backgrounds at the FCC-hh with $\sqrt{s}=100$ TeV. The distributions are for $f_{M,i}/\Lambda^4$
with $i=0,1,2,3,4,5,7$ , $f_{T,j}/\Lambda^4$  with $j=0,1,2,5,6,7$ and various backgrounds
for leptonic decay channel of the $W$-boson. In these figures we have taken
$f_{M,i}/\Lambda^4 = 100\hspace{1mm}{\rm TeV}^{-4}= 1\times 10^{-10}\hspace{1mm}{\rm GeV}^{-4}$
and $f_{T,j}/\Lambda^4 = 1\hspace{1mm}{\rm TeV}^{-4} = 1\times 10^{-12} \hspace{1mm}{\rm GeV}^{-4}$.}
  \label{Fig.4}
\end{figure}

\begin{figure}[ht]
  \begin{subfigure}[b]{0.5\linewidth}
    \centering
    \includegraphics[width=1.00\linewidth]{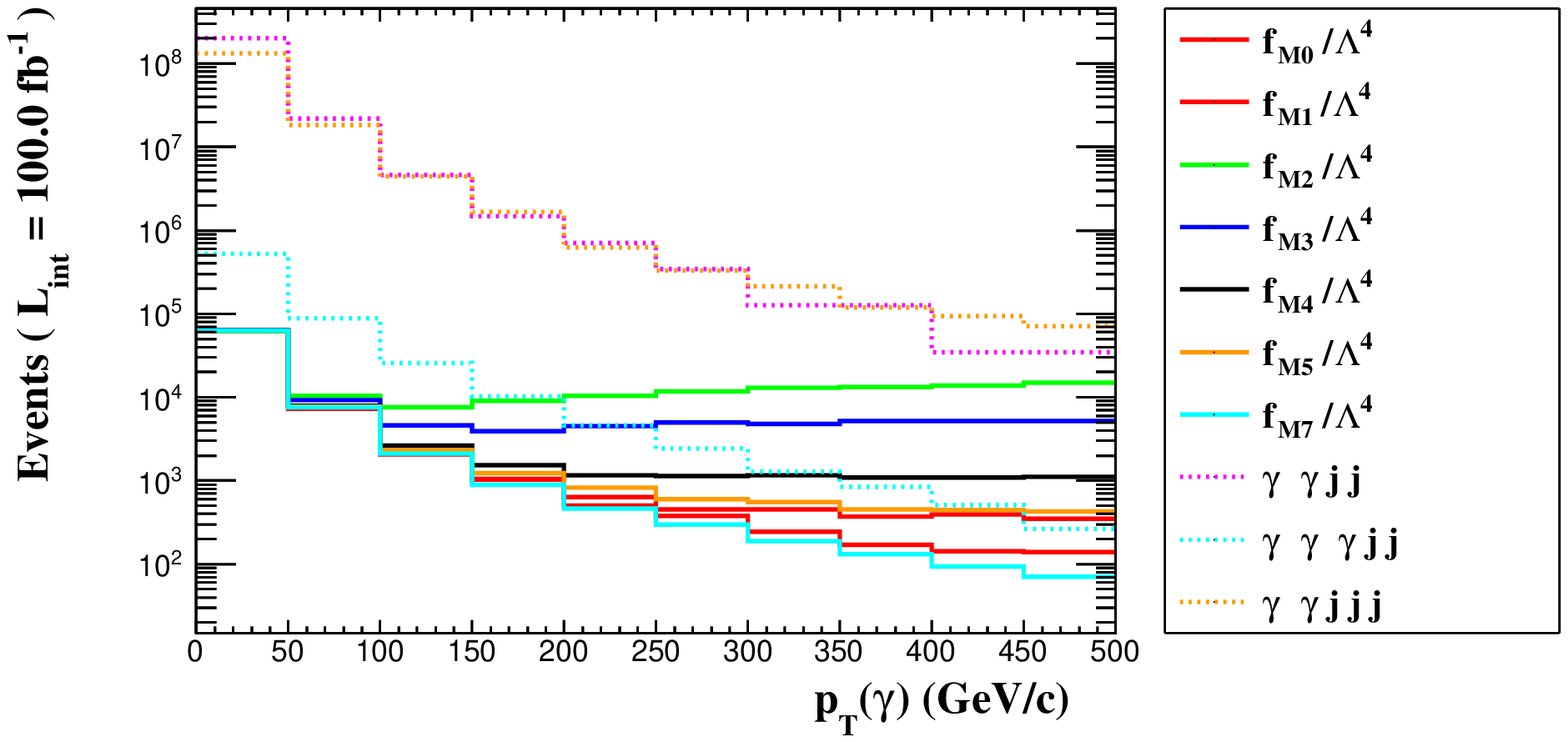}
    \caption{Hadronic decay}
    \label{fig8:a}
    \vspace{4ex}
  \end{subfigure}%%
  \begin{subfigure}[b]{0.5\linewidth}
    \centering
    \includegraphics[width=1.00\linewidth]{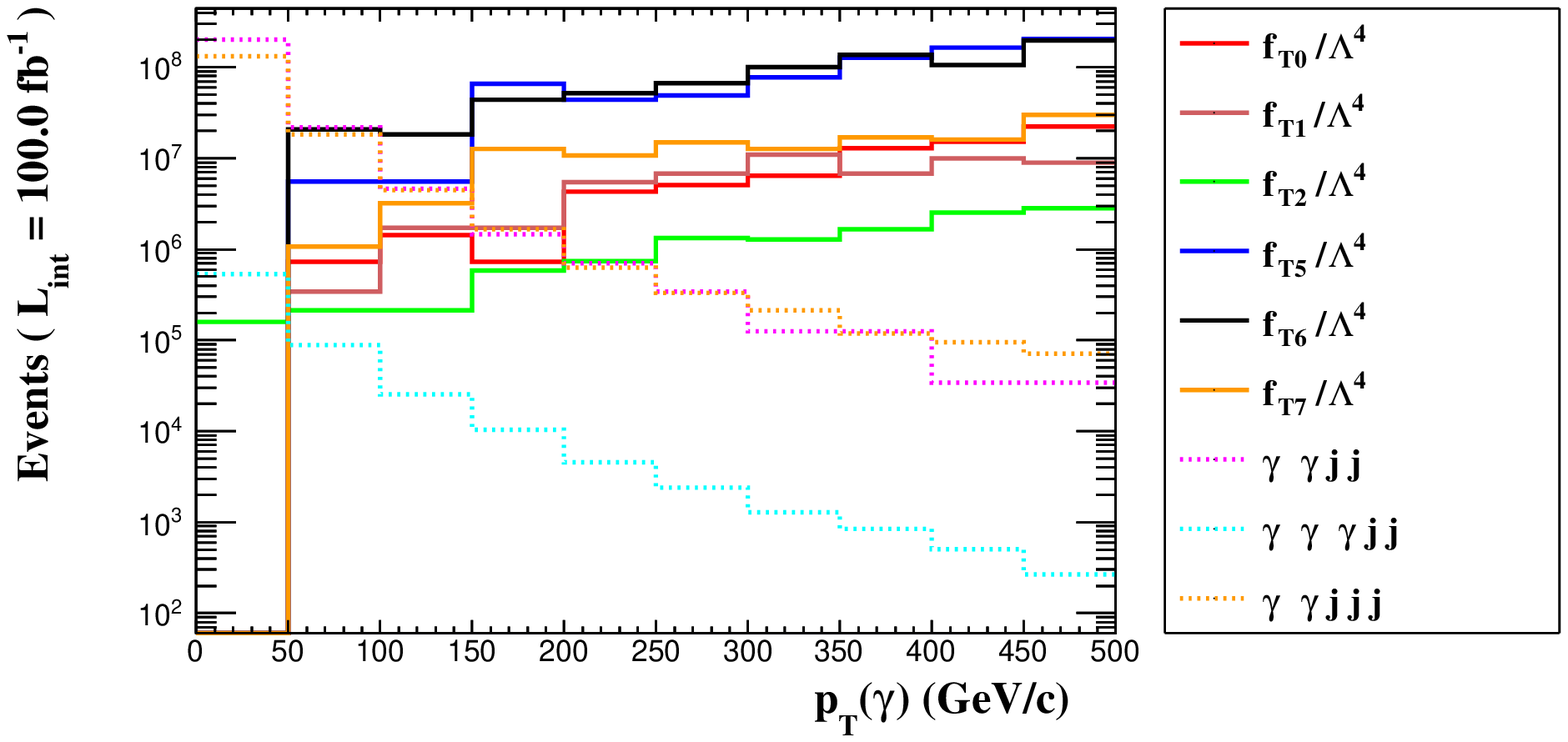}
    \caption{Hadronic decay}
    \label{fig8:b}
    \vspace{4ex}
  \end{subfigure}
  \caption{Same as in Fig. 5, but for hadronic decay. In these figures we have taken
$f_{M,i}/\Lambda^4 = 1\times 10^{-9}\hspace{1mm}{\rm GeV}^{-4}$
and $f_{T,j}/\Lambda^4 = 1\times 10^{-9} \hspace{1mm}{\rm GeV}^{-4}$.}
  \label{Fig.4}
\end{figure}


\begin{thebibliography}{99}


\bibitem{HL-LHC-HE-LHC} P. Azzi, {\it et al.}, {\it Standard Model Physics at the HL-LHC and HE-LHC}, arXiv:1902.04070v3 [hep-ph].

\bibitem{FCC-hh1} FCC web site http://cern.ch/fcc; also see M. Benedikt and F. Zimmermann, CERN Courier 2014,
                  http://cerncourier.com/cws/article/cern/56603.

\bibitem{FCC-hh2} Physics at the FCC-hh, a 100 TeV pp collider, CERN Yellow Reports: Monographs Volume 3/2017, CERN-2017-003-M,
                  Editor: M. L. Mangano.

\bibitem{FCC-hh3} A. Abada, {\it et al.},  {\it Eur. Phys. J. Special Topics} {\bf 228}, 755 (2019).

\bibitem{FCChe} Oliver Br\"{u}ning, John Jowett, Max Klein, Dario Pellegrini, Daniel Schulte and Frank Zimmermann,
                EDMS 17979910 FCC-ACC-RPT-0012, V1.0, 6 April, 2017. https://fcc.web.cern.ch/Documents/FCCheBaselineParameters.pdf.

\bibitem{Link-FCC-he-CERN} FCC Conceptual Design Report volumes 1-4: http://fcc-cdr.web.cern.ch

\bibitem{ILC-Brau} J. Brau, {\it et al.}, The International Linear Collider (ILC): A global project,
                   ${\rm https://ilchome.web.cern.ch/sites/ilchome.web.cern.ch/files/ILC_European_Strategy_Document-{ILCGeneral}.pdf (2018)}$.

\bibitem{CLIC-Burrows} P. N. Burrows, {\it et al.}, The Compact Linear Collider (CLIC)-2018 Summary Report,
                       {\bf CERN Yellow Rep.Monogr. 1802 (2018) 1-98}, arXiv:1812.06018 [physics.acc-ph].

\bibitem{CEPC-Ahmad} M. Ahmad, {\it et al.}, (The CEPC-SPPC Study Group), CEPC conceptual design report:
                     Volume 2-Physics $\&$ Detector, arXiv:1811.10545 [hep-ex].

\bibitem{TLEP-Bicer} M. Bicer, {\it et al.}, [TLEP design study working group collaboration].
                      First look at the physics case of TLEP. {\it JHEP} {\bf 01}, 164 (2014)

\bibitem{ATLAS-PRL2015} G. Aad, {\it et al.}, [ATLAS Collaboration], {\it Phys. Rev. Lett.} {\bf 115}, 031802 (2015).

\bibitem{ATLAS-PRD2016} G. Aad, {\it et al.}, [ATLAS Collaboration], {\it Phys. Rev.} {\bf D93}, 112002 (2016).

\bibitem{CMS-2105.12780} A. Tumasyan, {\it et al.}, [The CMS Collaboration], arXiv:2105.12780v2 [hep-ex].

\bibitem{PLB811-2020} A. M. Sirunyan, {\it et al.}, [The CMS Collaboration], {\it Phys.Lett.} {\bf B811}, 135988 (2020).

\bibitem{ALEPH-Barate} R. Barate, {\it et al.}, [ALEPH Collaboration], {\it Phys. Lett.} {\bf B462}, 389 (1999).

\bibitem{DELPHI-Abreu} P. Abreu, {\it et al.}, [DELPHI Collaboration], {\it Phys. Lett.} {\bf B459}, 382 (1999).

\bibitem{L3-Acciarri} M. Acciarri, {\it et al.}, [L3 Collaboration], {\it Phys. Lett.} {\bf B467}, 171 (1999).

\bibitem{OPAL-Abbiendi} G. Abbiendi, {\it et al.}, [OPAL Collaboration], {\it Eur. Phys. J.} {\bf C8}, 191 (1999).

\bibitem{CDF-Gounder} K. Gounder, [CDF Collaboration], hep-ex/9903038.

\bibitem{D0-Abbott} B. Abbott, {\it et al.}, [DØ Collaboration], {\it Phys. Rev.} {\bf D62}, 052005 (2000).

\bibitem{CMS-Chatrchyan} S. Chatrchyan, {\it etal.}, [CMS Collaboration], {\it JHEP} {\bf 07}, 116 (2013).

\bibitem{ATLAS-Aaboud} M. Aaboud, {\it et al.}, [ATLAS Collaboration], {\it Eur. Phys. J.} {\bf C77}, 646 (2017).

\bibitem{Stirling} W. J. Stirling and A. Werthenbach, {\it Eur. Phys. J.} {\bf C14}, 103 (2000).

\bibitem{Leil} G. A. Leil and W. J. Stirling, {\it J. Phys.} {\bf G21}, 517 (1995).

\bibitem{Bervan} P. J. Dervan, A. Signer, W.J. Stirling, and A. Werthenbach, {\it J. Phys.} {\bf G26}, 607 (2000).

\bibitem{Chong} C. Chong, {\it et al.}, {\it Eur. Phys. J.} {\bf C74}, 3166 (2014).

\bibitem{Koksal} M. Koksal and A. Senol, {\it Int. J. Mod. Phys.} {\bf A30}, 1550107 (2015).

\bibitem{Gutierrez} A. Guti\'errez-Rodr\'iguez, C. G. Honorato, J. Monta\~no and M. A. P\'erez, {\it Phys. Rev.} {\bf D89}, 034003 (2014).

\bibitem{Belanger} G. Belanger, F. Boudjema, Y. Kurihara, D. Perret-Gallix, and A. Semenov, {\it Eur. Phys. J.} {\bf C13}, 283 (2000).

\bibitem{Stirling1} W. J. Stirling and A. Werthenbach, {\it Phys. Lett.} {\bf B466}, 369 (1999).

\bibitem{Atag} S. Atag and I. Sahin, {\it Phys. Rev.} {\bf D75}, 073003 (2007).

\bibitem{Eboli} O. J. P. Eboli, M. C. Gonzalez-Garcia, and S. F. Novaes, {\it Nucl. Phys.} {\bf B411}, 381 (1994).

\bibitem{Eboli1} O. J. P. Eboli, M. B. Magro, P. G. Mercadante, and S. F. Novaes, {\it Phys. Rev.} {\bf D52}, 15 (1995).

\bibitem{Sahin} I. Sahin, {\it J. Phys.} {\bf G36}, 075007 (2009).

\bibitem{Koksal1} M. Koksal, V. Ari and A. Senol, {\it Adv. High Energy Phys.} {\bf 2016}, 8672391 (2016).

\bibitem{Koksal2} M. Koksal, {\it Mod. Phys. Lett.} {\bf A29}, 1450184 (2014).

\bibitem{Yang} D. Yang, Y. Mao, Q. Li, S. Liu, Z. Xu, and K. Ye, {\it JHEP} {\bf 1304}, 108 (2013).

\bibitem{Eboli2} O. J. P. Eboli, M. C. Gonzalez-Garcia, and S. M. Lietti, S. F. Novaes, {\it Phys. Rev.} {\bf D63}, 075008 (2001).

\bibitem{Bell} P. J. Bell, {\it Eur. Phys. J.} {\bf C64}, 25 (2009).

\bibitem{Ahmadov} A. I. Ahmadov, arXiv:1806.03460.

\bibitem{Schonherr} M. Schonherr, {\it JHEP} {\bf 1807}, 076 (2018).

\bibitem{Wen} Y. Wen, {\it et al.}, {\it JHEP} {\bf 1503}, 025 (2015).

\bibitem{Ye} K. Ye, D. Yang and Q. Li, {\it Phys. Rev.} {\bf D88}, 015023 (2013).

\bibitem{Eboli4} O. J. P. Eboli, M. C. Gonzalez-Garcia, and S. M. Lietti, {\it Phys. Rev.} {\bf D69}, 095005 (2004).

\bibitem{Perez} G. Perez, M. Sekulla and D. Zeppenfeld, {\it Eur. Phys. J.} {\bf C78}, 759 (2018).

\bibitem{Sahin1} I. Sahin and B. Sahin, {\it Phys. Rev.} {\bf D86}, 115001 (2012).

\bibitem{Baldenegro} C. Baldenegro, {\it et al.}, {\it JHEP} {\bf 1706}, 142 (2017).

\bibitem{Fichet} S. Fichet, {\it et al.}, {\it JHEP} {\bf 1502}, 165 (2015).

\bibitem{Pierzchala} T. Pierzchala and K. Piotrzkowski, {\it Nucl. Phys. Proc. Suppl.} {\bf 257}, 179 (2008).

\bibitem{LHeC-FCC-he-WWgg-Ari1} V. Ari, E. Gurkanli, A. A. Billur and M. K\"oksal, {\it Nucl. Phys.} {\bf B957}, 115102 (2020);
                                arXiv:1812.07187 [hep-ph].

\bibitem{LHeC-FCC-he-WWgg-Ari2} V. Ari, E. Gurkanli, A. Guti\'errez-Rodr\'iguez, M. A. Hern\'andez-Ru\'iz
                                and M. K\"oksal, {\it Eur. Phys. J. Plus} {\bf 135}, 336 (2020), arXiv:1911.03993 [hep-ph].

\bibitem{Eboli3} O. J. P. Eboli, M. C. Gonzalez-Garcia and J. K. Mizukoshi, {\it Phys. Rev.} {\bf D74}, 073005 (2006).

\bibitem{Chapon} E. Chapon, C. Royon and O. Kepka, {\it Phys. Rev.} {\bf D81}, 074003 (2010).

\bibitem{Senol} A. Senol and M. Koksal, {\it JHEP} {\bf 1503}, 139 (2015).

\bibitem{Koksal3} M. Koksal, {\it Eur. Phys. J. Plus} {\bf 130}, 75 (2015).

\bibitem{Senol1} A. Senol and M. Koksal, {\it Phys. Lett.} {\bf B742}, 143-148 (2015).

\bibitem{Aaboud} M. Aaboud, {\it et al.}, [ATLAS Collaboration], {\it Eur. Phys. J.} {\bf C77}, 646 (2017).

\bibitem{Gutierrez-EPJC81-2021} A. Guti\'errez-Rodr\'iguez, M. A. Hern\'andez-Ru\'iz, E. Gurkanli, V. Ari and M. Koksal,
                                {\it Eur. Phys. J.} {\bf C81}, 210 (2021).

\bibitem{twiki.cern} ${\rm https://twiki.cern.ch/twiki/bin/view/CMSPublic/PhysicsResultsSMPaTGC \sharp Figure\_7}$
${\rm \_limits\_on\_dimension\_8\_m}$

\bibitem{Eboli-PRD101-2020} E. da Silva Almeida, O. J .P. \'Eboli, M. C. Gonzalez-Garcia, {\it Phys. Rev.} {\bf D101}, 113003 (2020);
                            arXiv:2004.05174 [hep-ph].

\bibitem{Degrande} C. Degrande, {\it et al.}, arXiv: 1309.7890.

\bibitem{Data2020} P. A. Zyla, {\it et al.}, (Particle Data Group), {\it Prog. Theor. Exp. Phys.} {\bf 2020}, 083C01 (2020).

\bibitem{Baak} M. Baak, {\it et al.}, Working group report: precision study of electroweak interactions, arXiv:1310.6708.

%\bibitem{Hagiwara-NPB1987} K. Hagiwara, K. Hikasa, R. D. Peccei, D. Zeppenfeld, {\it Nucl. Phys.} {\bf B282}, 253 (1987).

%\bibitem{Baur-PLB1988} U. Baur and D. Zeppenfeld, {\it Phys. Lett.} {\bf B201}, 383 (1988).

%\bibitem{Hagiwara-PLB1992} K. Hagiwara, S. Ishihara, R. Szalapski, and D. Zeppenfeld, {\it Phys. Lett.} {\bf B283}, 353 (1992),

%\bibitem{Hagiwara-PRD1993} K. Hagiwara, S. Ishihara, R. Szalapski, and D. Zeppenfeld, {\it Phys. Rev.} {\bf D48}, 2182 (1993).

\bibitem{Ellison-ARNPS1998} J. Ellison and J. Wudka, {\it Annu. Rev. Nucl. Part. Sci.} {\bf 48}, 33 (1998).

%\bibitem{Weiglein-PR2006} G. Weiglein, {\it et al.}, (LHC/ILC Study Group), {\it Phys. Rep.} {\bf 426}, 47 (2006).

\bibitem{MadGraph} J. Alwall, M. Herquet, F. Maltoni, O. Mattelaer and T. Stelzer, {\it JHEP} {\bf 06}, 128 (2011).

\bibitem{AAlloul} A. Alloul, N. D. Christensen, C. Degrande, C. Duhr and B. Fuks, {\it Comput. Phys. Commun.} {\bf 185}, 2250 (2014),
                 arXiv:1310.1921 [hep-ph].

\bibitem{CDegrande} C. Degrande, C. Duhr, B. Fuks, D. Grellscheid, O. Mattelaer and T. Reiter, {\it Comput. Phys. Commun.} {\bf 183}, 1201 (2012),
                   arXiv:1108.2040 [hep-ph].

\bibitem{PDF-JHEP07-2002} J. Pumplin, D. R. Stump, J. Huston, H. L. Lai, P. Nadolsky and W. K. Tung, {\it JHEP} {\bf 07}, 012 (2002).

\bibitem{EPJC80-2020} A. Yilmaz, A. Senol, H. Denizli, I. Turk Cakir and O. Cakir, {\it Eur. Phys. J.} {\bf C80}, 173 (2020).

\bibitem{CMS-PAS-SMP-20-016} The CMS Collaboration, {\it Measurement of the electroweak production of $Z\gamma$ and two jets
                             in proton-proton collisions at $\sqrt{s} = 13$ TeV and constraints on dimension 8 operators},
                             CMS PAS SMP-20-016.

\bibitem{CMS-PAS-SMP-19-013} The CMS Collaboration, {\it Measurements of the $pp \to W^{\pm}\gamma\gamma$ and $pp \to Z\gamma\gamma$ cross
                             sections and limits on anomalous quartic gauge couplings at $\sqrt{s} = 13$ TeV}, CMS PAS SMP-19-013.

\bibitem{CMS-JHEP06-2020} A. M. Sirunyan, {\it et al.}, (CMS Collaboration), {\it JHEP} {\bf 06}, 076 (2020).

\bibitem{PRD83-2011} G. Bozzi, F. Campanario, M. Rauch, and D. Zeppenfeld, {\it Phys. Rev.} {\bf D83}, 114035 (2011).

\bibitem{Koksal6} M. K\"oksal, A. A. Billur, A. Guti\'errez-Rodr\'iguez and M. A. Hern\'andez-Ru\'iz,
                  {\it Phys. Lett.} {\bf B808}, 135661 (2020), arXiv:1910.06747 [hep-ph].

\bibitem{Gutierrez2} A. Guti\'errez-Rodr\'iguez, M. K\"oksal, A. A. Billur and M. A. Hern\'andez-Ru\'iz,
                     {\it J. Phys.} {\bf G47}, 055005 (2020), arXiv:1910.02307 [hep-ph].

\bibitem{Billur5} A. A. Billur, M. K\"oksal, A. Guti\'errez-Rodr\'iguez and M. A. Hern\'andez-Ru\'iz, arXiv:1909.10299 [hep-ph].

\bibitem{Koksal7} M. K\"oksal, A. A. Billur, A. Guti\'errez-Rodr\'iguez and M. A. Hern\'andez-Ru\'iz,
                  {\it Int. J. Mod. Phys.} {\bf A35}, 2050178 (2020), arXiv:1905.02564 [hep-ph].

\bibitem{Gutierrez3} A. Guti\'errez-Rodr\'iguez, M. K\"oksal, A. A. Billur and M. A. Hern\'andez-Ru\'iz, arXiv:1903.04135 [hep-ph].

\bibitem{CMS-JHEP08} V. Khachatryan, {\it et al.}, (CMS Collaboration), {\it JHEP} {\bf 08}, 119 (2016).

\bibitem{CMS-JHEP06-2017} V. Khachatryan, {\it et al.}, (CMS Collaboration), {\it JHEP} {\bf 06}, 106 (2017).

\bibitem{ATLAS-PRD94-2016} M. Aaboud, {\it et al.} (ATLAS Collaboration), {\it Phys. Rev.} {\bf D94}, 032011 (2016).

\bibitem{OPAL-PRD70} G. Abbiendi, {\it et al.}, [OPAL Collaboration], {\it Phys. Rev.} {\bf D70}, 032005 (2004).

\bibitem{OPAL-PLB580} G. Abbiendi, {\it et al.}, [OPAL Collaboration], {\it Phys. Lett.} {\bf B580}, 17 (2004).

\bibitem{DELPHI-EPJC31} DELPHI Collaboration, {\it Eur. Phys. J.} {\bf C31}, 139 (2003).

\bibitem{L3-PLB527} P. Achard, {\it et al.}, [L3 Collaboration], {\it Phys. Lett.} {\bf B527}, 29 (2002).

\bibitem{D0-PRD88} V. M. Abazov, {\it et al.}, [D0 Collaboration], {\it Phys. Rev.} {\bf D88}, 012005 (2013).

\bibitem{CMS-EPJC73} The CMS Collaboration, {\it Eur. Phys. J.} {\bf C73}, 2283 (2013).

\bibitem{CMS-JHEP10} The CMS Collaboration, {\it J. High Energy Phys.} {\bf 10}, 072 (2017).

\bibitem{ATLAS-PRD94} The ATLAS Collaboration, {\it Phys. Rev.} {\bf D 94}, 032011 (2016).

\end{thebibliography}
\end{document}